\pdfoutput=1
\documentclass{JINST}

\usepackage{array}
\usepackage{lineno}
\usepackage{afterpage}
\usepackage{subfigure}

\newcolumntype{L}{>{\centering\arraybackslash}m{2.5cm}}
\newcolumntype{S}{>{\centering\arraybackslash}m{1.5cm}}
\newcolumntype{V}{>{\centering\arraybackslash}m{3.5cm}}
\newcolumntype{W}{>{\centering\arraybackslash}m{0.8cm}}
\newcolumntype{Y}{>{\centering\arraybackslash}m{1.50cm}}
\title{Detection of low energy antiproton annihilations in a segmented silicon detector}

\author{S. Aghion$^{a,b}$, O. Ahl\'{e}n$^c$, A. S. Belov$^d$, G. Bonomi$^{e,f}$,P. Br\"{a}unig$^g$, J. Bremer$^c$, R. S. Brusa$^h$, G. Burghart$^c$, L. Cabaret$^i$, M. Caccia$^j$, C. Canali$^k$, R. Caravita$^l$, F. Castelli$^l$, G. Cerchiari$^m$, S. Cialdi$^l$, D. Comparat$^i$, G. Consolati$^{a,b}$, J.H. Derking$^c$, S. Di Domizio$^n$, L. Di Noto$^h$, M. Doser$^c$, A. Dudarev$^c$, R. Ferragut$^{a,b}$, A. Fontana$^f$, P. Genova$^f$, M. Giammarchi$^b$, A. Gligorova$^o$\thanks{Corresponding author.}~, S. N. Gninenko$^d$, S. Haider$^c$, J. Harasimowicz$^p$, T. Huse$^q$, E. Jordan$^m$, L. V. J\o rgensen$^c$, T. Kaltenbacher$^c$, A. Kellerbauer$^m$, A. Knecht$^c$, D. Krasnick\'{y}$^r$, V. Lagomarsino$^r$, A. Magnani$^{f,s}$, S. Mariazzi$^t$, V. A. Matveev$^{d,u}$, F. Moia$^{a,b}$, G. Nebbia$^v$, P. N\'{e}d\'{e}lec$^w$, N. Pacifico$^o$, V. Petr\'{a}\v{c}ek$^x$, F. Prelz$^b$, M. Prevedelli$^y$, C. Regenfus$^k$, C. Riccardi$^{s,f}$, O. R\o hne$^q$, A. Rotondi$^{s,f}$, H. Sandaker$^o$, A. Sosa$^p$, M. A. Subieta Vasquez$^{e,f}$, M. \v{S}pa\v{c}ek$^x$, G. Testera$^n$, C. P. Welsch$^p$ and S. Zavatarelli$^n$, \linebreak\vspace{1.5 cm}
{\linebreak}(AEgIS collaboration) \\
\llap{$^a$}Politecnico di Milano, 
  Piazza Leonardo da Vinci 32, 20133 Milano, Italy\\
\llap{$^b$}Istituto Nazionale di Fisica Nucleare,
  Sez. di Milano, Via Celoria 16, 20133 Milano, Italy\\
\llap{$^c$}European Organisation for Nuclear Research, Physics Department,
 1211 Geneva 23, Switzerland\\
\llap{$^d$}Institute for Nuclear Research of the Russian Academy of Sciences,
 Moscow 117312, Russia\\
\llap{$^e$}University of Brescia, Department of Mechanical and Industrial Engineering, Via Branze 38, 25133 Brescia, Italy\\
\llap{$^f$}Istituto Nazionale di Fisica Nucleare, Sez. di Pavia,
 Via Agostino Bassi 6, 27100 Pavia, Italy\\
\llap{$^g$}University of Heidelberg, Kirchhoff Institute for Physics,
  Im Neuen heimer Feld 227, 69120 Heidelberg, Germany\\
\llap{$^h$}Department of Physics, University of Trento and TIFPA-INFN, Via Sommarive 14, 38123    Povo, Trento,Italy \\
\llap{$^i$}Laboratoire Aim\'{e} Cotton, CNRS, Universit\'{e} Paris Sud, ENS Cachan, B\^{a}timent 505, Campus d'Orsay, 91405 Orsay Cedex, France\\
\llap{$^j$}Insubria University,
  Como-Varese, Italy\\
\llap{$^k$}University of Zurich, Physics Institute,
  Winterthurerstrasse 190, 8057 Zurich, Switzerland\\
\llap{$^l$}University of Milano, Department of Physics,
  Via Celoria 16, 20133 Milano, Italy\\
\llap{$^m$}Max Planck Institute for Nuclear Physics,
  Saupfercheckweg 1, 69117 Heidelberg, Germany\\
\llap{$^n$}Istituto Nazionale di Fisica Nucleare, Sez. di Genova,
  Via Dodecaneso 33, 16146 Genova, Italy\\
\llap{$^o$}University of Bergen, Institute of Physics and Technology,
  All\'{e}gaten 55, 5007 Bergen, Norway\\
\llap{$^p$}University of Liverpool and the Cockroft Institute, Liverpool, Sci-Tech Daresbury,
  Keckwick Lane, Daresbury, Warrington, WA4 4AD, United Kingdom\\
\llap{$^q$}University of Oslo, Department of Physics,
  Sem S\ae landsvei 24, 0371 Oslo, Norway\\
\llap{$^r$}University of Genoa, Department of Physics,
  Via Dodecaneso 33, 16146 Genova, Italy\\
\llap{$^s$}University of Pavia, Department of Nuclear and Theoretical Physics,
  Via Bassi 6, 27100 Pavia, Italy\\
\llap{$^t$}Stefan Meyer Institute for subatomic Physics, Boltzmanngasse 3, 1090 Vienna, Austria\\
\llap{$^u$}Joint Institute for Nuclear Research,
  141980 Dubna, Russia\\
\llap{$^v$}Istituto Nazionale di Fisica Nucleare, Sez. di Padova,
  Via Marzolo 8, 35131 Padova, Italy\\
\llap{$^w$}Claude Bernard University Lyon 1, Institut de Physique Nucl\'{e}aire de Lyon, 4 Rue Enrico Fermi, 69622 Villeurbanne, France\\
\llap{$^x$}Czech Technical University in Prague, FNSPE,
  B\v{r}ehov\'{a} 7, 11519 Praha 1, Czech Republic\\
\llap{$^y$}University of Bologna, Department of Physics,
  Via Irnerio 46, 40126 Bologna, Italy\\

  E-mail: \email{Angela.Gligorova@cern.ch}
}                                          

\abstract{The goal of the AE$\mathrm{\bar{g}}$IS experiment at the Antiproton Decelerator (AD) at CERN, is to measure directly the Earth's gravitational acceleration on antimatter by measuring the free fall of a pulsed, cold antihydrogen beam.
The final position of the falling antihydrogen will be detected by a position sensitive detector.
This detector  will consist  of an active silicon part, where the annihilations take place, followed by an emulsion part. Together, they allow to achieve 1$\%$ precision on the measurement of $\bar{g}$ with about 600 reconstructed and time tagged annihilations.

We present here the prospects for the development of the AE$\mathrm{\bar{g}}$IS silicon position sentive detector and the results from the first beam tests on a monolithic silicon pixel sensor, along with a comparison to Monte Carlo simulations.}


\begin{document}
\linenumbers
\clearpage
\section{Introduction}
\begin{figure*}[b]
\begin{center}
\includegraphics[width=15cm]{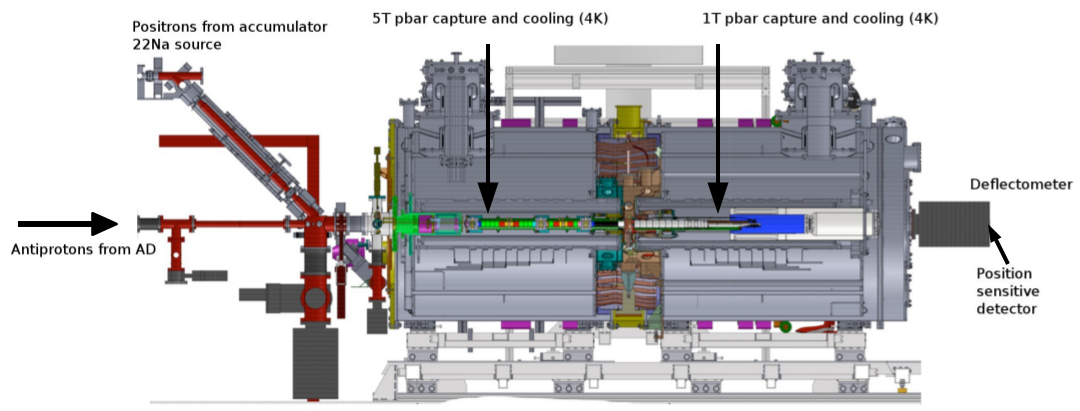}   
\caption{ Schematic view of the central region of the AE$\mathrm{\bar{g}}$IS experiment. }
\label{fig:AEgIS-all}
\end{center}
\end{figure*}

The AE$\mathrm{\bar{g}}$IS experiment \cite{Aegis2007} at CERN (fig. \ref{fig:AEgIS-all}) aims at verifying the Weak Equivalence Principle for antimatter by measuring the Earth's gravitational acceleration \textit{g} for antihydrogen. Several attempts have been made in the past to measure the gravitational constant for antimatter, both for charged \cite{ps200, nieto1991} and neutral antiparticles \cite{Brando1981,sn1987a, Mills2002}. However, none of these experiments brought to conclusive results. Recently, a study from the ALPHA collaboration \cite{alphag} sets limits to the ratio of gravitational mass to the inertial mass of antimatter but is yet far from testing the equivalence principle. Another experiment, GBAR, \cite{gbar} has been proposed but not yet built.

Cold antihydrogen (100 mK) in Rydberg states will be produced through the charge exchange reaction between Rydberg positronium and cold antiprotons stored in a Penning trap \cite{doser2012}. Applying an appropriate electric field will accelerate the formed antihydrogen in a horizontal beam, with a typical axial velocity distribution spanning a few 100 m/s~\cite{testera2008}.

Some of the trajectories will be selected through a moir\'{e} deflectometer \cite{moire}, which will consist of two vertical gratings producing a fringe pattern on a downstream annihilation plane (see fig. \ref{fig:moiresilicon}). This plane will be the first layer of the position sensitive detector where the antihydrogen will impinge with energies of the order of meV and annihilate. The vertical deflection of the pattern is proportional to the gravitational constant to be measured. Over a flight path of $\sim1$ m, the deflection is expected in the order of $\sim20$ $\mathrm{\mu}$m for a 1 g vertical acceleration \cite{Aegis2007}. A vertical resolution better than 10 $\mathrm{\mu}$m is required to meet the goal of 1$\%$ precision on the $\bar{g}$ measurement with 600 reconstructed and time tagged annihilations~\cite{Amsler2012}.

According to the current design, the position sensitive detector will be a hybrid detector consisting of an active silicon part, where the annihilation takes place, followed by an emulsion part \cite{Amsler2012, Emulsion2013}. 
The silicon detector will provide online measurement and diagnostics of the antiproton annihilations as well as the necessary time of flight information. 

The aim of the present study is to perform the first measurement and direct detection of slow antiproton ($\sim$ few 100 keV) annihilations in silicon.
This is the first step towards understanding the signature of antihydrogen annihilations, which is one of the most fundamental aspects of designing a silicon position sensitive detector for AE$\mathrm{\bar{g}}$IS. To our knowledge, only in one other experiment annihilations in a silicon sensor were directly detected and simulated \cite{McGaughey1986}. However, much faster antiprotons were used in that study (608 MeV/c) than in the study presented here.

\section{Development of the silicon detector for AE$\mathrm{\bar{g}}$IS}

In AE$\mathrm{\bar{g}}$IS, the silicon detector will act as the annihilation surface. Kinetic energy of the antihydrogen atom will be insufficient to generate a detectable signal, so the antihydrogen will be indirectly detected through the detection of the annihilation products. We will now present available experimental data on the annihilation process of antihydrogen (antiprotons) in matter and the available Monte Carlo tools for its simulation. This constitutes the basis for the design of the AE$\mathrm{\bar{g}}$IS silicon detector, which will be presented in \ref{aegissidet}.

\subsection{Annihilation of antiprotons in silicon}

\label{sec:detection}

The annihilation process of antihydrogen in matter is similar to the one of an antiproton as the positron annihilates immediately when meeting an atomic electron. 
Previous experiments at LEAR \cite{Bendiscioli1994} have studied annihilations of antiprotons in elements with different Z. In this process, the antiproton loses energy as it traverses matter and annihilates with a proton at rest creating charged ($1.53\pm0.03$ per annihilation per charge sign) and neutral pions ($1.96\pm0.23$ per annihilation). For elements with atomic numbers $>$1 the average ratio is shifted towards producing more negatively charged pions, due to the possible annihilation of the antiproton with nuclear neutrons. The pions produced in the annihilation may further interact with other nucleons resulting in nuclear fragments and isolated neutrons and protons.
For silicon, the stopping power of the lowest incoming antiproton energy so far measured (0.188 MeV) shows to be 32$\%$ lower than for  protons  \cite{StoppingPower}. 

Antimatter annihilation has been detected with silicon sensors previously \cite{Athena2004}, through the detection of pions emitted in the annihilation process. However, in our present application, for the first time the antiproton annihilates with a nucleon in the bulk of the detector itself. These pions are \emph{Minimum Ionizing Particles} (MIPs) depositing $\sim0.3$ keV/$\mathrm{{\mu}m}$ \cite{Spieler2005} in matter, a negligible fraction compared with their average momentum of $\sim350$ MeV/c \cite{asterix85}.

\begin{figure*}[htp]
\centering
\begin{minipage}{.5\textwidth}
  \centering
\includegraphics[width=8cm]{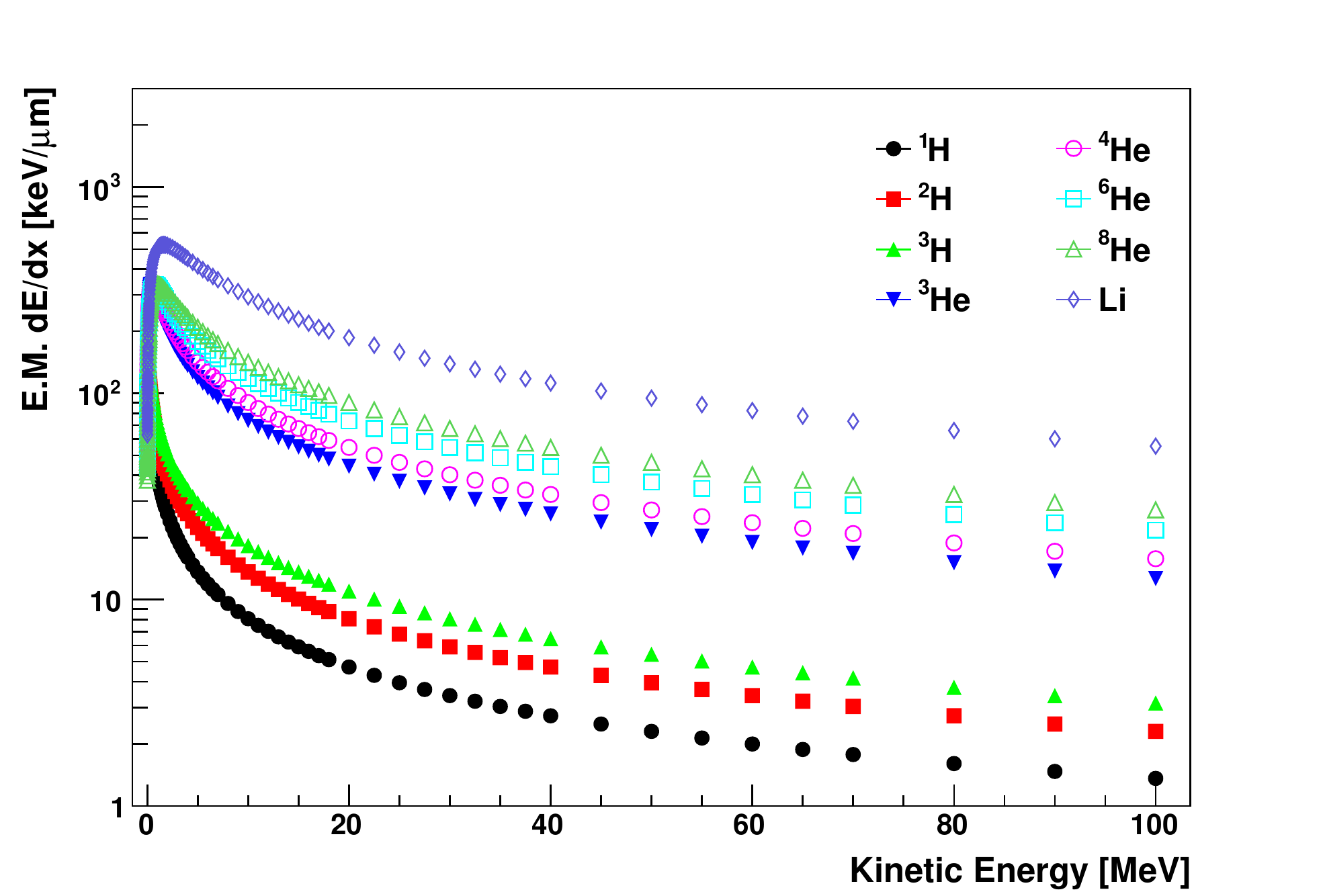}
\caption{Energy deposition in silicon for different nuclear fragments that can be generated in an annihilation event, calculated with the SRIM package \cite{SRIM}.}
\label{fig:EnergyDeposition}
\end{minipage}%
\hspace*{1cm}
\begin{minipage}{.5\textwidth}
  \centering
\includegraphics[width=8cm]{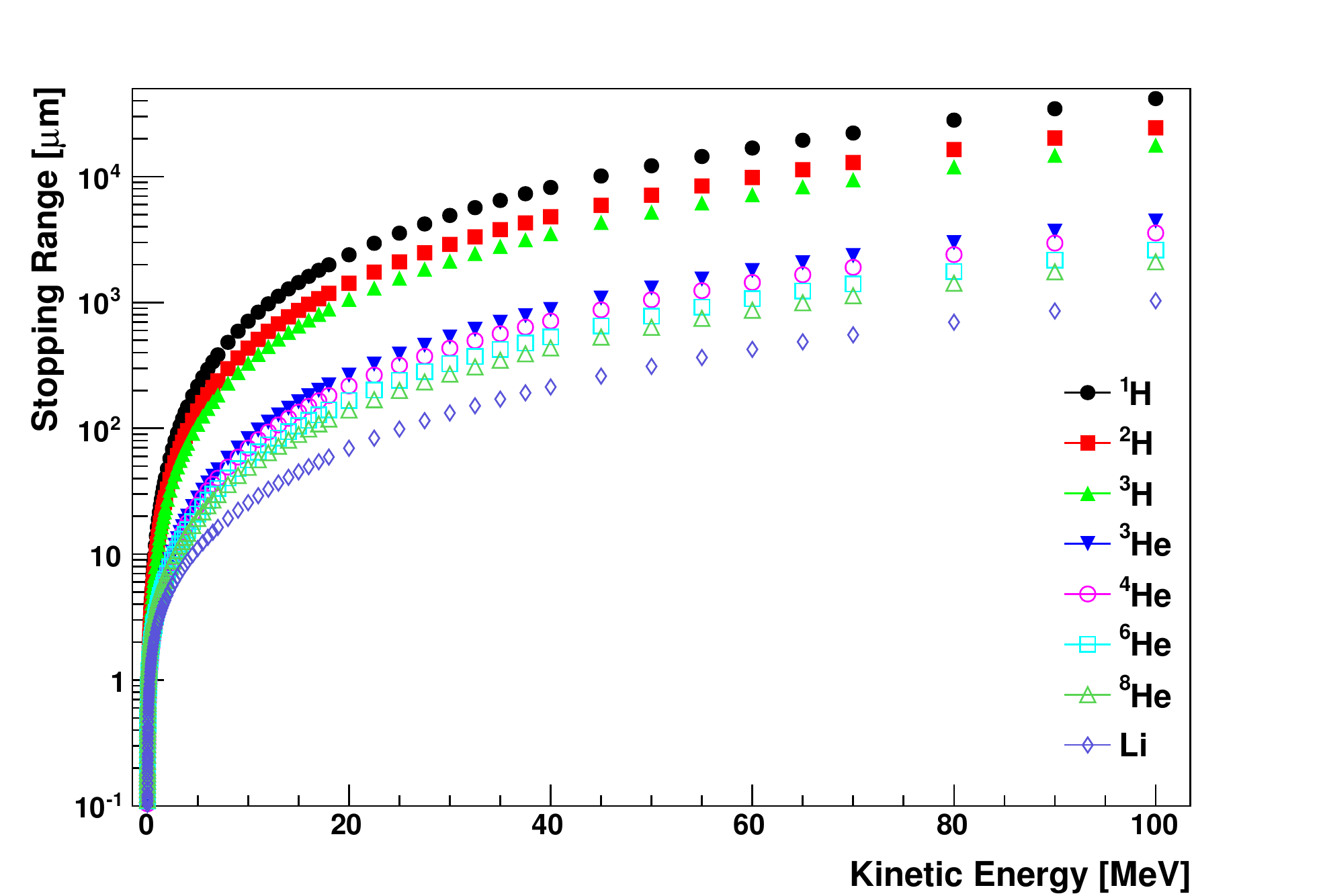}
\caption{Stopping range in silicon for different nuclear fragments that can be generated in an annihilation event, calculated with the SRIM package.}
\label{fig:StoppingRange}
\end{minipage}
\end{figure*}

When the annihilation takes place on-sensor, the largest fraction of deposited energy is due to the heavy fragments. These fragments are \emph{Highly Ionizing Particles} (or HIPs).
Energy deposits and ranges in silicon for different annihilation products simulated using the SRIM \cite{SRIM} package are shown in fig. \ref{fig:EnergyDeposition} and \ref{fig:StoppingRange}. HIPs (slow protons and heavier ions) deposit locally (within a few or tens of $\mathrm{\mu}$m from the interaction point) all of their kinetic energy. 
It becomes thus evident that being able to discriminate between the signal produced by HIPs or MIPs in the detector, can help increasing significantly the resolution on the annihilation position.

\begin{figure}[h!]
\begin{center}
\includegraphics[width=8cm]{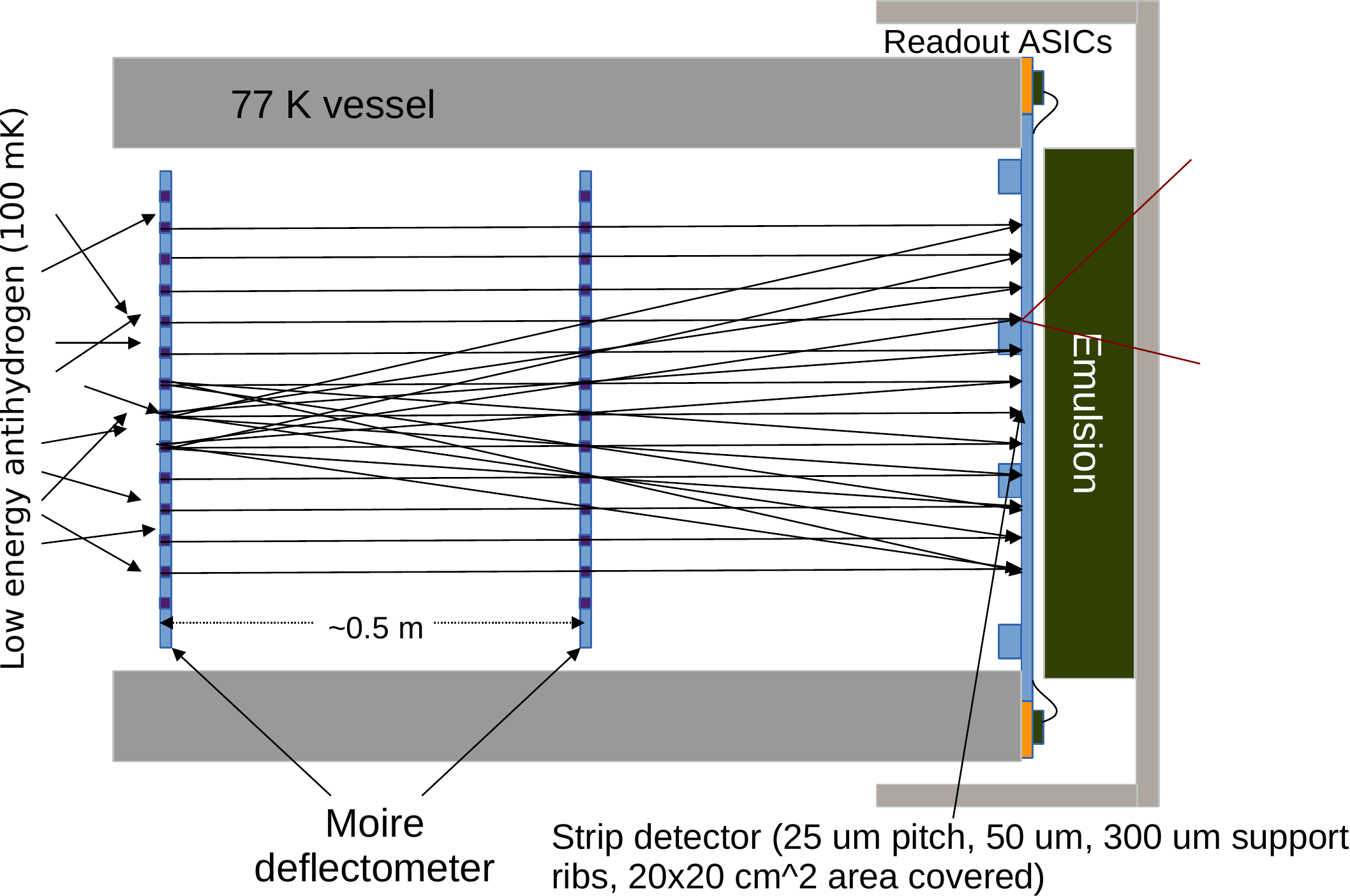} 
\caption{The moir\'{e} deflectometer producing a pattern on the position sensitive detector, where more particle paths intersect at the detector plane.}
\label{fig:moiresilicon}
\end{center}
\end{figure}

\subsection {Monte Carlo simulations}
\label{Montecarlo}

In the present work we compare data with Monte Carlo simulations, using GEANT4, release 4.9.5.p01, interfaced with VMC (Virtual Monte Carlo) software, release v2-13c \cite{Hrivnacova}. Two particular GEANT4 models were studied, CHIPS (QGSP BERT) and FTFP (FTFP BERT TRV).

The CHIPS (CHiral Invariant Phase Space) model \cite{Degtyarenko} is a 3D quark-level event generator for the fragmentation of excited hadronic systems into individual hadrons, whereas the FTFP model \cite{Galoyan} relies on a string model to describe the interactions between quarks. 

\begin{figure}
\begin{center}
\includegraphics[width=8cm]{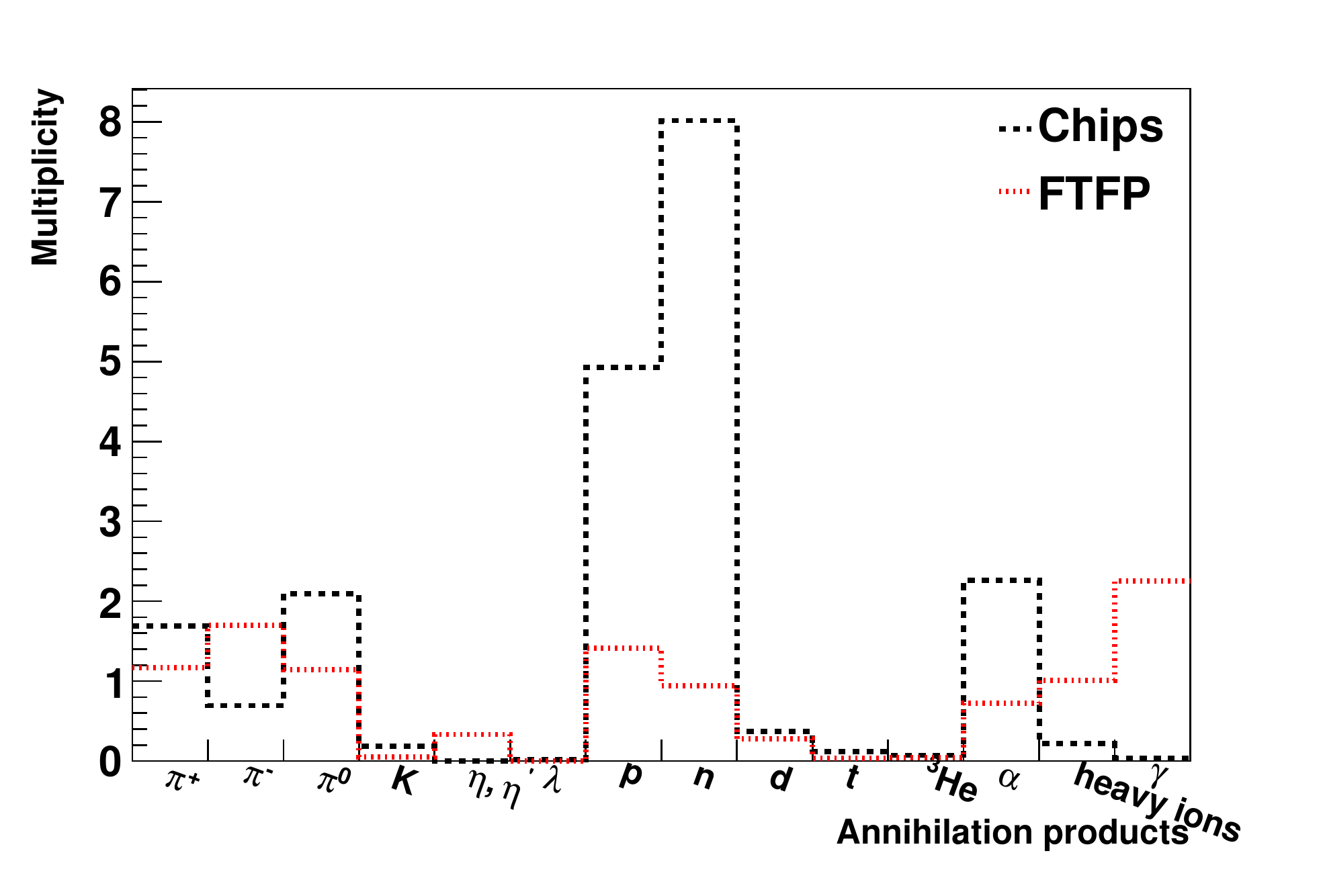}
\caption{Multiplicity of different annihilation products (per annihilation) as predicted by the two models CHIPS and FTFP,  over the whole kinetic energy spectrum.}
\label{fig:Prongs}
\end{center}
\end{figure}

The CHIPS and FTFP models differ in the production rate and in the composition of the annihilation products. CHIPS produces heavy nuclear fragments in only 20 $\%$ of the events while FTFP generates heavy fragments in all of them. In addition, CHIPS produces more than three times as many protons, neutrons and alpha particles in each collision, as seen in fig. \ref{fig:Prongs}, which provides the multiplicities for the different products for annihilations at rest.

Both models can simulate annihilation of antiprotons with nuclei, though comparison of simulations to data for low-energy antiprotons in silicon is missing.
CHIPS simulations have been previously compared with uranium and carbon data, while the newer FTFP still lacks comparison to data for antiproton energies below 120 MeV \cite{Geant4pl}.

Table \ref{tab:yields} shows a comparison of experimental values obtained for  $^{12}\mathrm{C}$ and $^{40}\mathrm{Ca}$, the two elements closest to silicon, with LEAR \cite{Markiel1988}, and the simulated values for the same elements and silicon. However, the values presented are for higher energies ($>$ 6 MeV) than in this study. The table shows that for the kinetic energy range of 6-18 MeV,  FTFP describes the data obtained for protons better than CHIPS. On the other hand, CHIPS describes better the experimental values for ion species with higher atomic numbers and for higher energies.

\begin{table*}[!htp]
\centering
\rotatebox{90}{
\footnotesize
\begin{tabular}{|c|Y|Y|Y|Y|Y|Y|Y|Y|Y|}
\hline
  		&Energy (MeV)	&LEAR $^{12}\mathrm{C}$		&CHIPS $^{12}\mathrm{C}$ 	& FTFP $^{12}\mathrm{C}$&LEAR $^{40}\mathrm{Ca}$	&CHIPS $^{40}\mathrm{Ca}$ 	& FTFP $^{40}\mathrm{Ca}$&CHIPS $^{28}\mathrm{Si}$	 & FTFP $^{28}\mathrm{Si}$			\\
\hline
p 		& 6-18 		& $23.3\pm2.0$ 			& $168.0\pm1.0$ 		& $56.0\pm0.8$		& $74.2\pm4.1$ 			& $172.0\pm1.0$ 		& $60.2\pm0.8$		 & $170.0\pm1.0$		 & $58.6\pm0.8$ 				\\
d 		& 8-24 		& $9.3\pm0.8$ 			& $15.9\pm0.4$ 			& $12.1\pm0.3$		& $18.1\pm1.1$ 			& $14.9\pm0.4$			& $12.0\pm0.3$		 & $15.9\pm0.4$ 		 & $12.2\pm0.3$ 				\\
t 		& 11-29 	& $4.5\pm0.4$ 			& $2.8\pm0.2$			& $1.3\pm0.1$		& $5.7\pm0.4$ 			& $2.7\pm0.2$			& $1.0\pm0.1$		 & $3.0\pm0.2$  		 & $1.5\pm0.1$  				\\
$^3\mathrm{He}$ & 36-70 	& $1.72\pm0.17$ 		& $0.19\pm0.01$ 		& $0.11\pm0.03$		& $2.22\pm0.17$ 		& $0.22\pm0.05$			& $0.13\pm0.04$		 & $0.23\pm0.05$		 & $0.16\pm0.04$				\\
$\alpha$ 	& 36-70 	& $1.14\pm0.12$ 		& $1.8\pm0.1$ 			& $0$			& $2.18\pm0.16$ 		& $1.9\pm0.1$	 		& $0$			 & $1.8\pm0.1$  		 & $0$  					\\
\hline
\end{tabular}
}
\caption{Measured and simulated production yields (for 100 annihilations) for the most important nuclear fragments produced in annihilations of antiprotons with high $A$ nuclei. Experimental data 
is from LEAR \cite{Markiel1988} for $^{12}\mathrm{C}$ and $^{40}\mathrm{Ca}$, the two elements closest to silicon. \emph{Energy} refers to the kinetic energy of the annihilations products. These measured values are compared with the simulated values for calcium, carbon and silicon using the two GEANT4 models, CHIPS and FTFP.  FTFP describes the data obtained with protons better than CHIPS, while CHIPS seems to be a better description for ion species with higher atomic numbers and higher energies.}
\label{tab:yields}
\end{table*}
\normalsize

\afterpage{\clearpage}

\subsection{Detector requirements and design}
\label{aegissidet}

As already shown in fig. \ref{fig:moiresilicon}, the AE$\mathrm{\bar{g}}$IS silicon position sensitive detector will act as a separation membrane between the ultra high vacuum of the antihydrogen formation and transport region and the secondary vacuum where the emulsion planes will be positioned. The resulting design includes an array of co-planar single-sided silicon strip sensors, built with a strip pitch of 25 $\mathrm{\mu}$m and mounted on a silicon mechanical support wafer, hosting the readout electronics. This system will provide with the one-dimensional vertical (y) deflection information, though an approach based on resistive strips, able to provide the x coordinate as well, as demonstrated in \cite{cnmresistive}, is currently under study.

A further requirement of the silicon detector will be a thickness, in the active regions, of 50 $\mathrm{\mu}$m. This will allow to minimize the scattering of annihilation products, detected further downstream by the emulsion detector, allowing for a precise vertex reconstruction. To achieve the goal, thick support ribs will guarantee the mechanical stability of the system, with size and position of the ribs being optimized as to allow for the maximum efficiency of the detector in areas were a higher beam luminosity is expected.

Finally, in order to avoid the black body radiation coming from the detector to increase the antiproton plasma temperature (which would increase the thermal velocity of the antihydrogen), the whole detector system will be kept at cryogenic temperatures (77 K or lower). This will require the electronics to be designed for such conditions. The feasibility of operation of standard CMOS readout ASICs in cryogenic temperatures has already been proven in \cite{Regenfus2003}. The ASIC design for AE$\mathrm{\bar{g}}$IS, under development, will rely on an improved integration and communication protocol (enabling the readout of $\sim$3000 strips) and a wider dynamic range, to cope with the high energy deposited in the sensor from the annihilation events.

Given the complex nature of the annihilation process, Monte Carlo simulations will be required to validate reconstruction algorithms to be implemented in the final system. Part of the aim of the present work is the validation of the available simulation physics model, in the particular case of direct annihilation in a silicon sensor, with data available for the first time for low antiproton energies.

\section{Test beam setup}
\label{sec:test facility}

\subsection{Antiproton source and test facility}
\label{ss:facility}

The AE$\mathrm{\bar{g}}$IS experiment is situated at the Antiproton Decelerator (AD) which delivers $\sim3 \times 10^{7}$ low energy (5.3 MeV) and bunched ($\sim$120 ns) antiprotons every $\sim100$ s. During  tests  in May 2012 the first section of the AE$\mathrm{\bar{g}}$IS experiment was in place, comprising a 5 T superconducting solenoid magnet enclosing a Penning trap in an ultrahigh vacuum (UHV) of  $10^{-11}$  mbar. 

\begin{figure*}[htp]
\centering
\includegraphics[width=15cm]{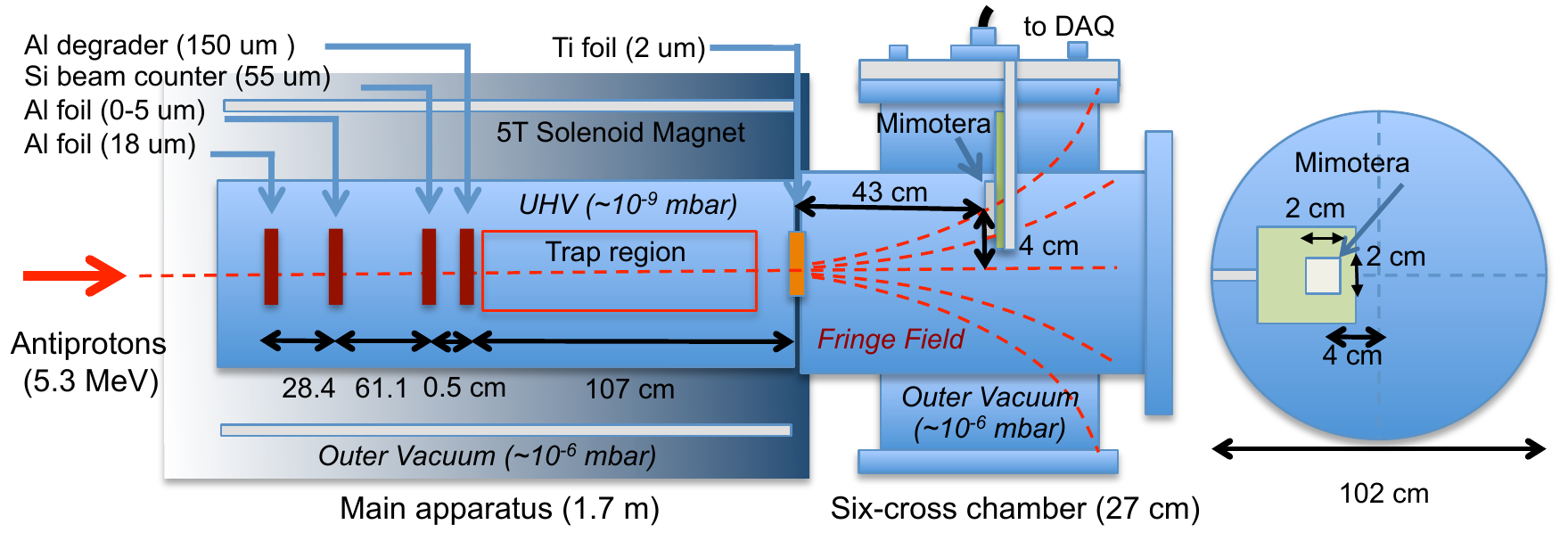} 
\caption{Top view (left) and axial view (right) of the test set-up. The center of the silicon detector (MIMOTERA) is installed 40 mm off axis and 430 mm from the main apparatus to avoid saturation due to the high beam intensity.}
\label{fig:Aegis2012}
\end{figure*}

While passing through the AE$\mathrm{\bar{g}}$IS apparatus, the antiprotons lose energy  first through two aluminum degraders, one fixed (18$\pm2.7$ $\mathrm{{\mu}}$m) and one mobile (0.8$\pm0.2$, 2$\pm0.5$, 3$\pm0.75$, 4$\pm1$ and 5 $\pm1.25$ $\mathrm{{\mu}}$m), then a silicon beam counter (55$\pm5.5$ $\mathrm{{\mu}}$m)\cite{BeamCounter} and another fixed aluminum degrader (150 $\pm15$ $\mathrm{{\mu}}$m) as shown in fig.~\ref{fig:Aegis2012}. After this, less than 1\% of the incoming antiprotons from the AD are trapped in flight by the Penning trap, while the rest continue downstream. 

\begin{figure*}[htp]
\centering
\includegraphics[width=7cm]{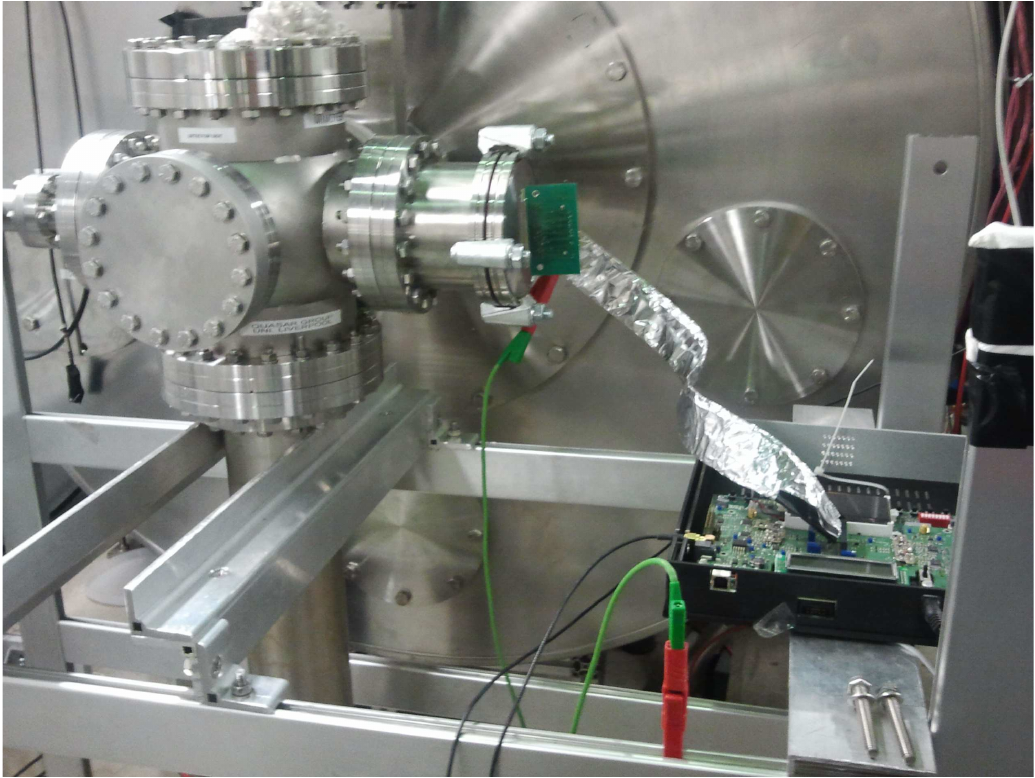} 
\caption{Photo of the six-way cross vacuum chamber in testbeam. The MIMOTERA is shown mounted on the  right hand flange together with its readout system.}
\label{fig:mimoterainstallation}
\end{figure*}

Before entering a six-cross vacuum chamber, where the detector was mounted (fig. \ref{fig:Aegis2012}) the antiproton beam traversed a 2 $\mathrm{{\mu}}$m thick titanium foil used to separate the UHV region from the secondary vacuum ($\sim10^{-7}$ mbar). In the six-way cross the antiprotons were deviated by the solenoid fringe field before hitting the silicon detector, which was mounted perpendicular to the beam and 40 mm off axis (fig. \ref{fig:Aegis2012} and \ref{fig:mimoterainstallation}).

To overcome the unavoidable small inaccuracies in the stopping power calculation  through the degraders' total thickness, the simulation (see sec. \ref{Montecarlo}) was independently tuned against the antiproton trapping efficiency during the tests of the antiproton capture trap. The simulated trapping efficiency with 229 $\mu$m of degrading material was equivalent to the real efficiency obtained with 225 $\mu$m of degraders. Nevertheless, the effect of both 225 and 229 $\mu$m silicon equivalent degrading material thicknesses were simulated and compared with data presented here for completeness. 

\begin{figure*}[htp]
\centering
\begin{minipage}{.5\textwidth}
\centering
\includegraphics[width=7cm]{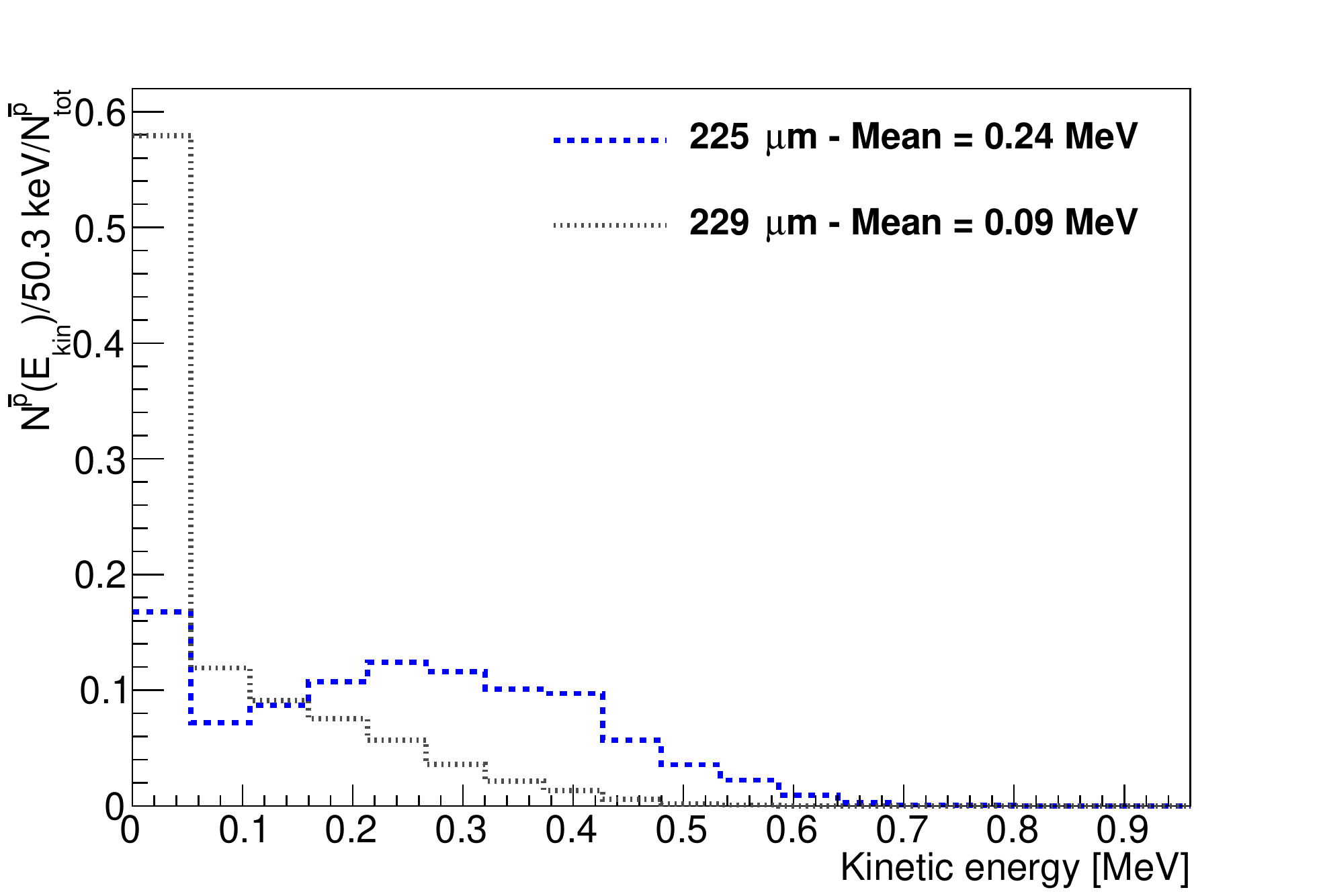} 
\caption{Kinetic energy distribution of the antiprotons before they reach the detector, as simulated with GEANT4.}
\label{fig:kinetic_energy}
\end{minipage}%
\hspace*{1cm}
\begin{minipage}{.5\textwidth}
\centering
\includegraphics[width=7cm]{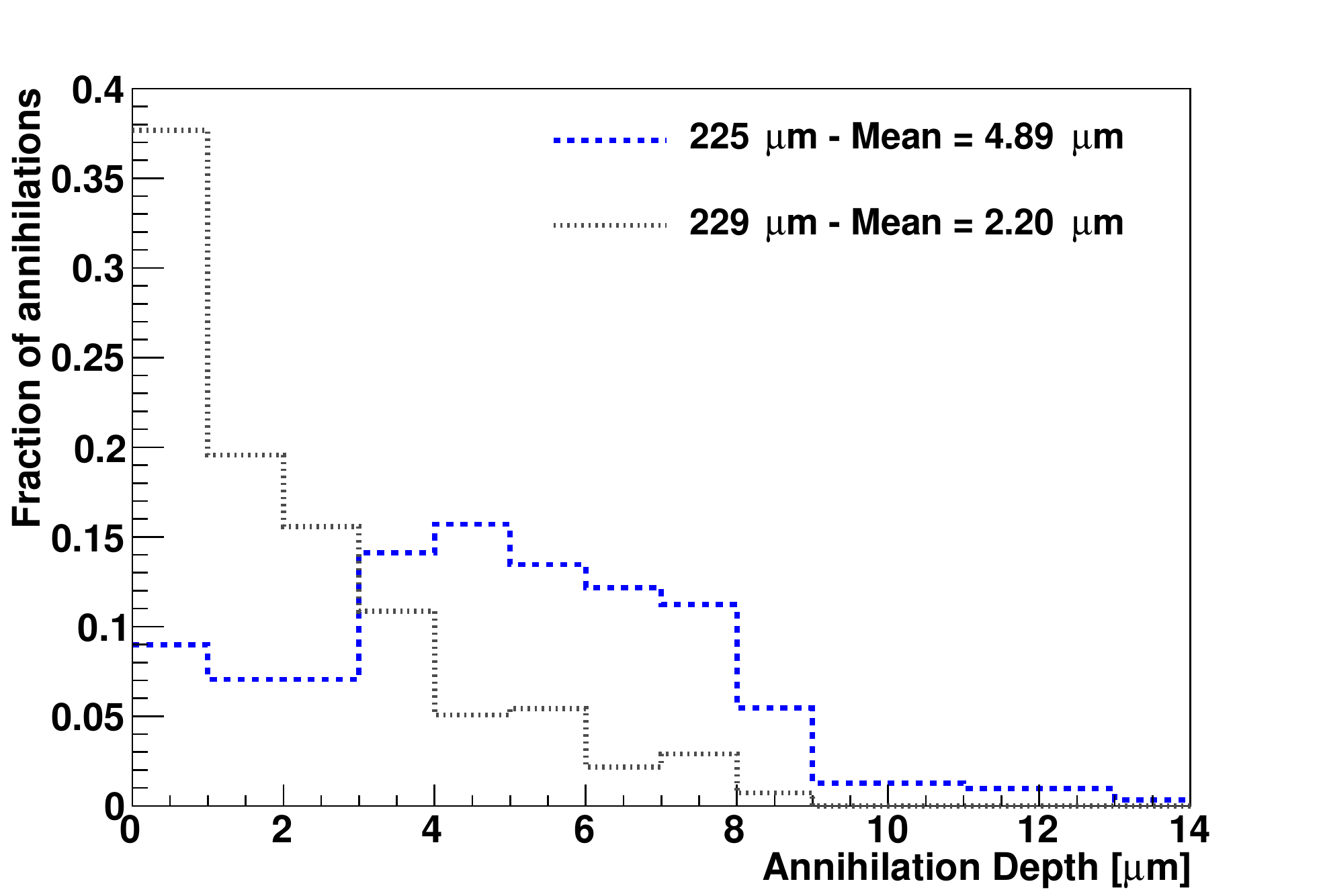} 
\caption{GEANT4 simulation showing the annihilation depth, as calculated from the kinetic energy distribution in fig. \protect\ref{fig:kinetic_energy}.}
\label{fig:kinetic_depths}
\end{minipage}
\end{figure*}

Fig. \ref{fig:kinetic_energy} shows the kinetic energy distribution of the antiprotons just before reaching the MIMOTERA detector as simulated with GEANT4. The average kinetic energy according to simulations was $\sim250$ keV for 225 ${\mu}\mathrm{m}$ material and $\sim100$ keV for 229 ${\mu}\mathrm{m}$. This energy is higher than the energy of the antihydrogen in the final system (meV), but much lower than any energy tested to date. The same simulation shows that $\sim60\%$ of the antiprotons coming from the AD reached the six-way cross chamber. The corresponding distribution of annihilation depths is shown in fig. \ref{fig:kinetic_depths}.

From the GEANT4 simulations (see sec. \ref{Montecarlo}) we could also estimate the spatial distribution of the antiproton beam. The resulting incident angle of antiprotons on the MIMOTERA was of 4.5$\pm$1.1$^\circ$ with respect to the normal to the detector plane.

In order to study the absorption effect on antiprotons and to verify them against the simulations, we covered 2/3 of the detector surface with three very thin aluminum foils (3, 6 and 9 ${\mu}\mathrm{m}$). The foils were suspended parallel to the detector surface at a distance $\sim5$ mm by means of three thin copper wires with a gauge of 300 $\mu$m, also running on the part not covered by the foils.

\subsection{The MIMOTERA detector}
\label{ss:mimoimplementation}

The MIMOTERA \cite{Boll2011} is a monolithic active pixel sensor in CMOS technology. It is characterized by a large area (17x17 $\mathrm{mm^{2}}$), a coarse granularity (with a square pixel of 153 $\mathrm{{\mu}}$m size) and a dynamic range over 3 orders of magnitude. Moreover, it is back-illuminated, with an entrance window $\sim$100 nm thick in addition to the 14 $\mathrm{{\mu}}$m thick sensitive layer. The detector has a global shutter and a continuous read-out with no dead-time: in AE$\mathrm{\bar{g}}$IS, impinging particles were identified by processing the difference between the frame containing the antiproton spill and the previous one (differential mode).

The MIMOTERA has been designed to be virtually unaffected by cross-talk, in virtue of the presence of multiple readout diodes for each pixel. More details can be found in \cite{Badano2005, BollThesis}.

The full well capacity of the pixels in the MIMOTERA corresponds to a deposited energy of $\sim$ 30 MeV/pixel.

\subsection{Calibration of the MIMOTERA detector and clustering}

The MIMOTERA was designed for the profilometry of radiotherapy beams applications for which no exact knowledge of the deposited energy is required. Therefore, to determine the amount of energy deposited in the detector, the response of the MIMOTERA was calibrated using a red laser source (${\lambda}=660$ nm).

The laser light, coming from a custom laser diode assembly at CERN, was directed by means of a fiber-coupled focuser onto the aperture window of the detector. A 5 ns  pulsed  signal was used to trigger both the laser diode and the MIMOTERA DAQ, which was operated at 2.5 MHz. 

To obtain an absolute value  for  the number of free carriers generated with the laser, the same laser was used to induce a transient charge pulse on a PAD diode, 300 ${\mu}$m thick, manufactured by HIP (Helsinki Institute for Physics) on Magnetic Czochralski silicon. All the light coming from the focuser was projected onto the optical window of the diode,  which had the same kind of passivation  layer as the MIMOTERA (100 nm $\mathrm{SiO_2}$). 

The signal, decoupled from the DC bias voltage by means of a Picosecond 5531 bias-tee, was read and acquired with a 500 MHz LeCroy oscilloscope. The unamplified signal was integrated up to $\sim100$ ns, where the transfer function of the electronics was measured to be constantly null. 

\begin{figure}
   \subfigure[]{
     \includegraphics[width=8cm]{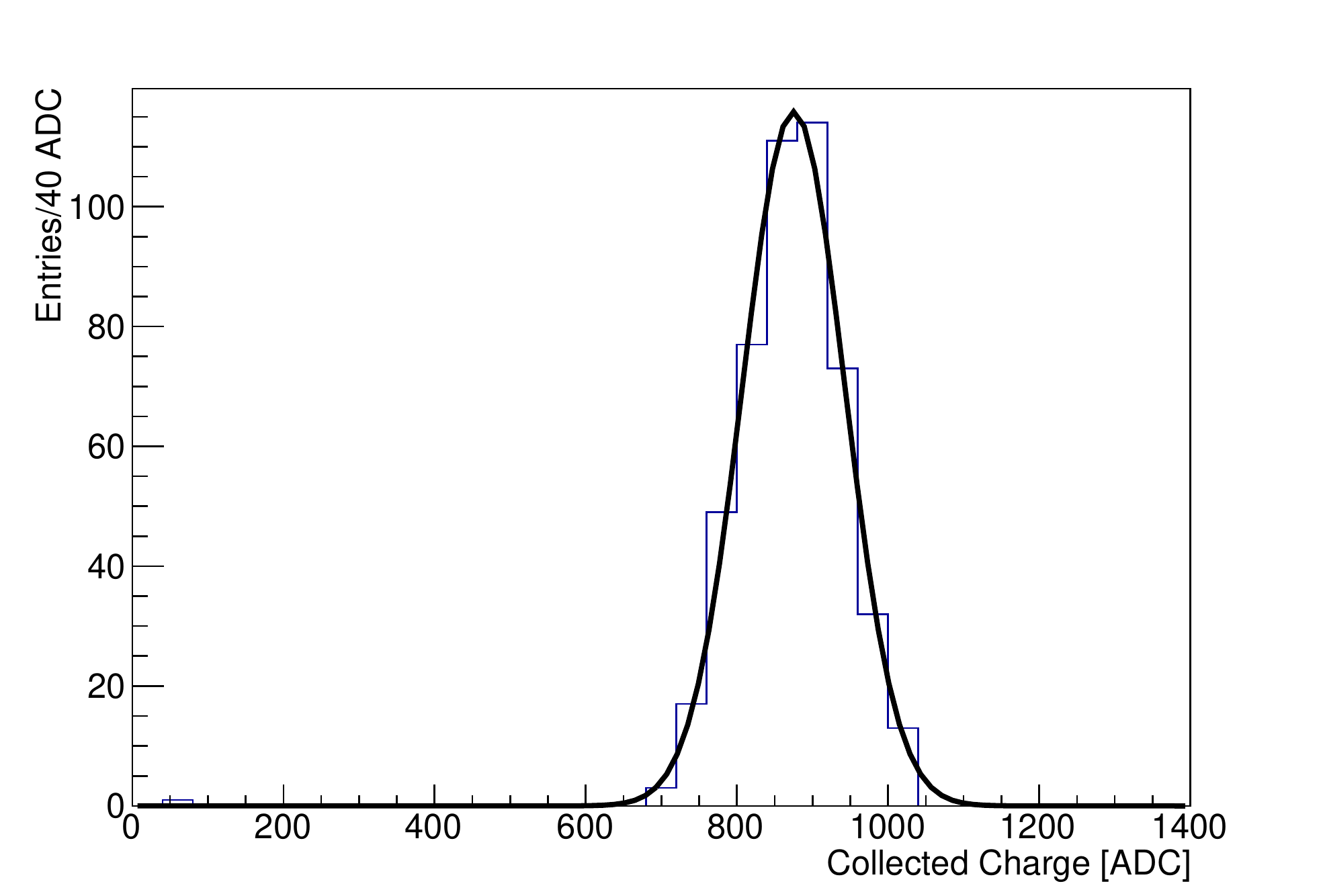}
     \label{fig:caliba}
   }~
   \subfigure[]{
     \includegraphics[width=8cm]{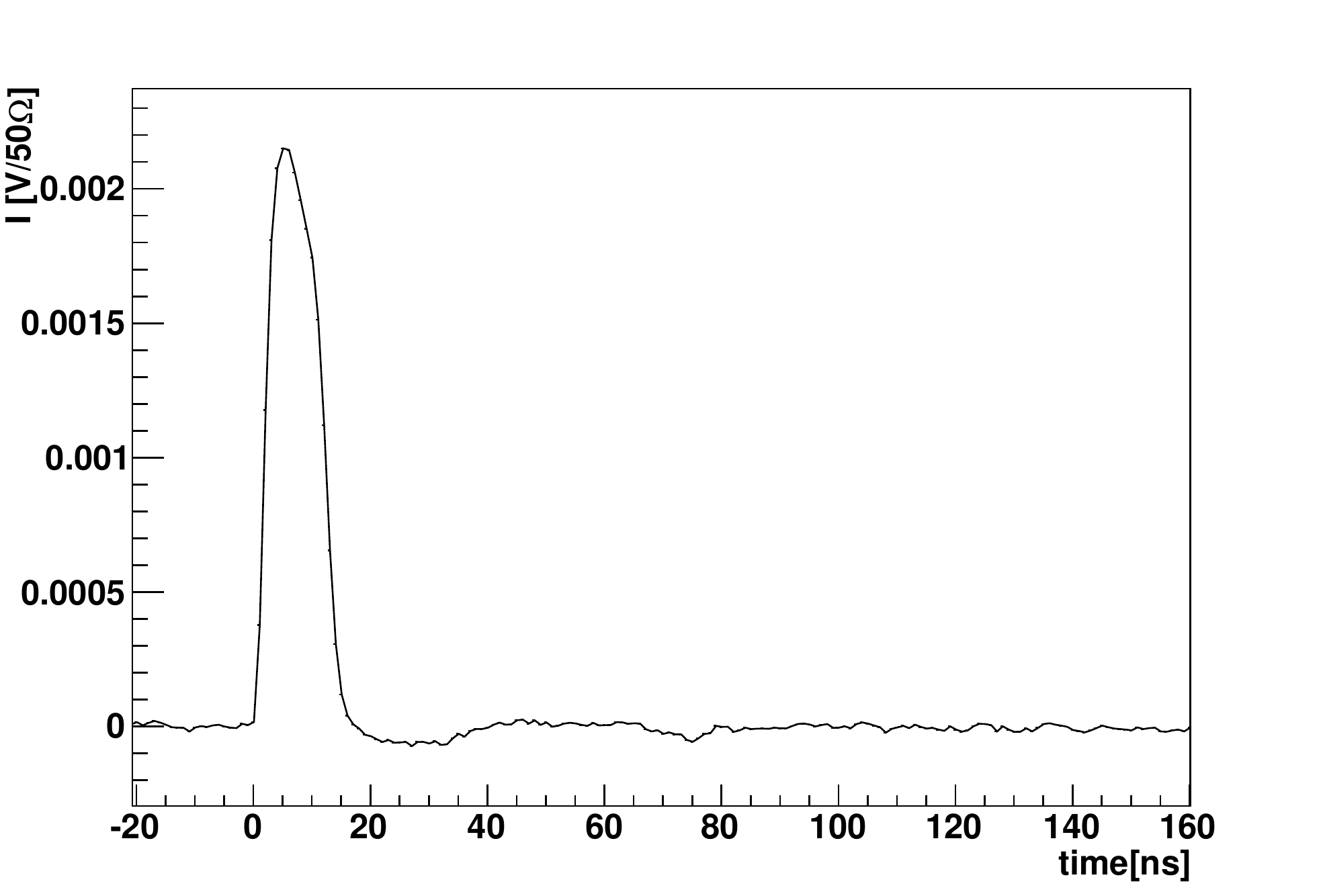}
     \label{fig:calibb}
   }
   \caption{Left: distribution of the signal generated by the laser in the MIMOTERA over 491 laser pulses, fitted with a gaussian curve. Right: Transient Current pulse from the HIP diode, as acquired through the oscilloscope, averaged over 1024 pulses.}
   \label{fig:calib}
\end{figure}

Fig. \ref{fig:calib} shows the signal distribution in ADC as acquired by the mimotera and a signal transient from the diode as induced in both cases by the laser beam.

Since the absorption length for 660 nm red light in silicon is $\sim3.3$ ${\mu}$m \cite{Palik85}, the thickness of the active region of the MIMOTERA detector allows to collect more than 98 $\%$ of the generated charge carriers. As the remaining 2 $\%$ could be either reflected or transimetted at the interface with the substrate, where the refraction index is unknown, the full 2 $\%$ systematic error was added to the calibration factor as a conservative estimate. By comparing the analog integrated pulse with the pixel charge digitally sampled by the MIMOTERA, we calculate a calibration factor of $(4889\pm100)\:\mathrm{eV/ADC\:count}$. A study verifying the linearity of the MIMOTERA detector can already be found in \cite{Boll2011}.

\begin{figure*}[htp]
\centering
\includegraphics[width=8cm]{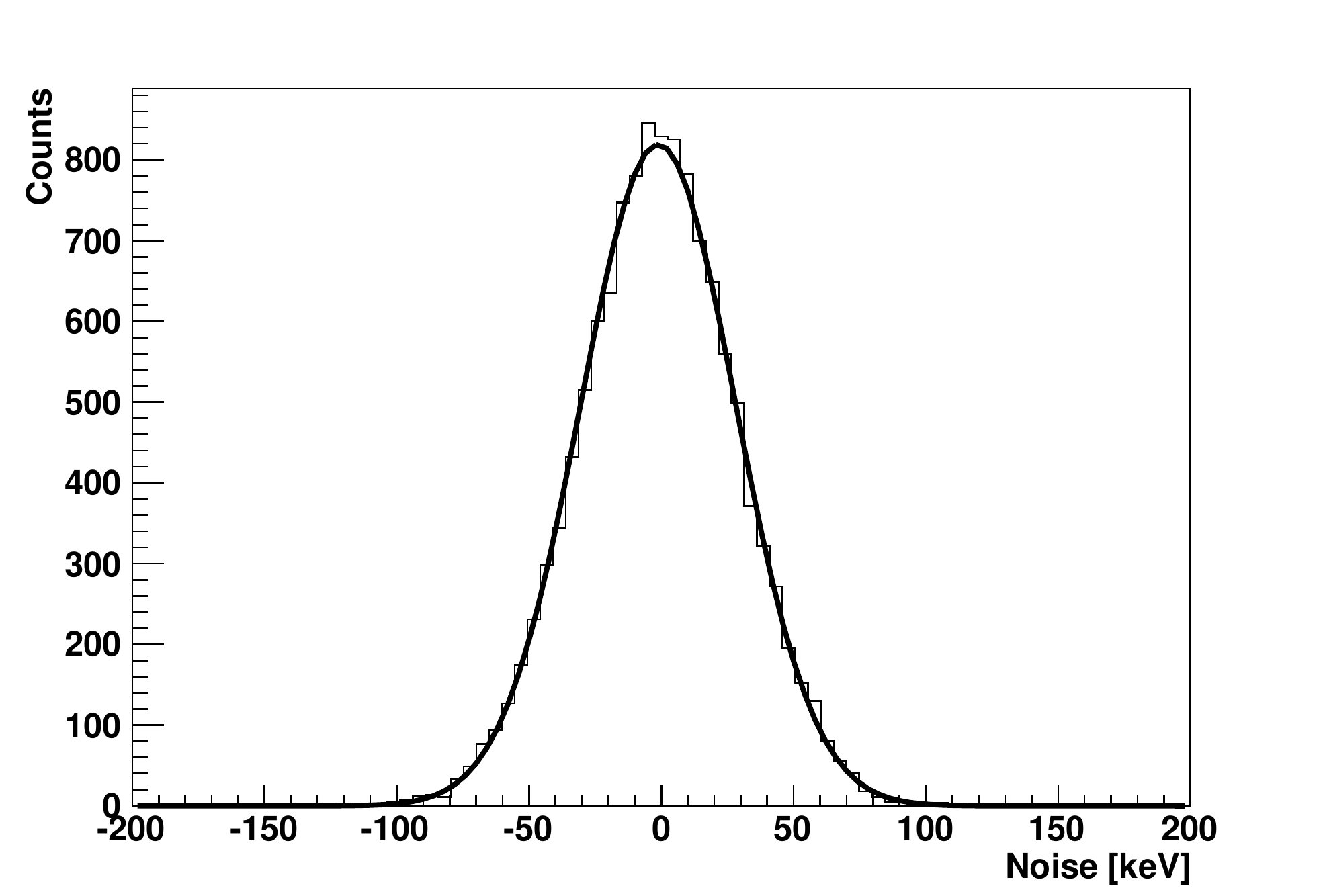} 
\caption{Noise spectrum of the MIMOTERA detector for one non-triggered frame.}
\label{fig:noise}
\end{figure*}

The single pixel noise in the experiment was measured to be 30.3 keV, with fairly low non-gaussian tails (fig. \ref{fig:noise}).


The single pixel energy distribution is shown in fig. \ref{fig:pixelenergy}, before and after subtracting the noise by fitting a Gaussian to the negative values (where there is no signal). The residual entries with energies lower than 5 noise RMS can be attributed to MIP-like pions (depositing between 4.2 keV to $\sim$ 65 keV depending on the crossing angle) and protons which for a wide energy range ($>$ 50 MeV) have a dE/dx$\simeq$2 keV/$\mu$m, (see fig. \ref{fig:EnergyDeposition}). This could possibly explain the peak observed at $\sim$30 keV. More detailed studies in this energy region will be performed in the future beam tests using detectors with higher sensitivity to low energies.

The complex nature of the annihilation process (see sec. \ref{sec:detection}) was not known and we had no estimation on how much of the energy would be deposited away from the annihilation point, for instance when a high energy particle creates a long track and deposits its energy in a Bragg peak several pixels away. However, having a thin detector would naturally reduce this contribution.

\begin{figure}
   \subfigure[]{
     \includegraphics[width=8cm]{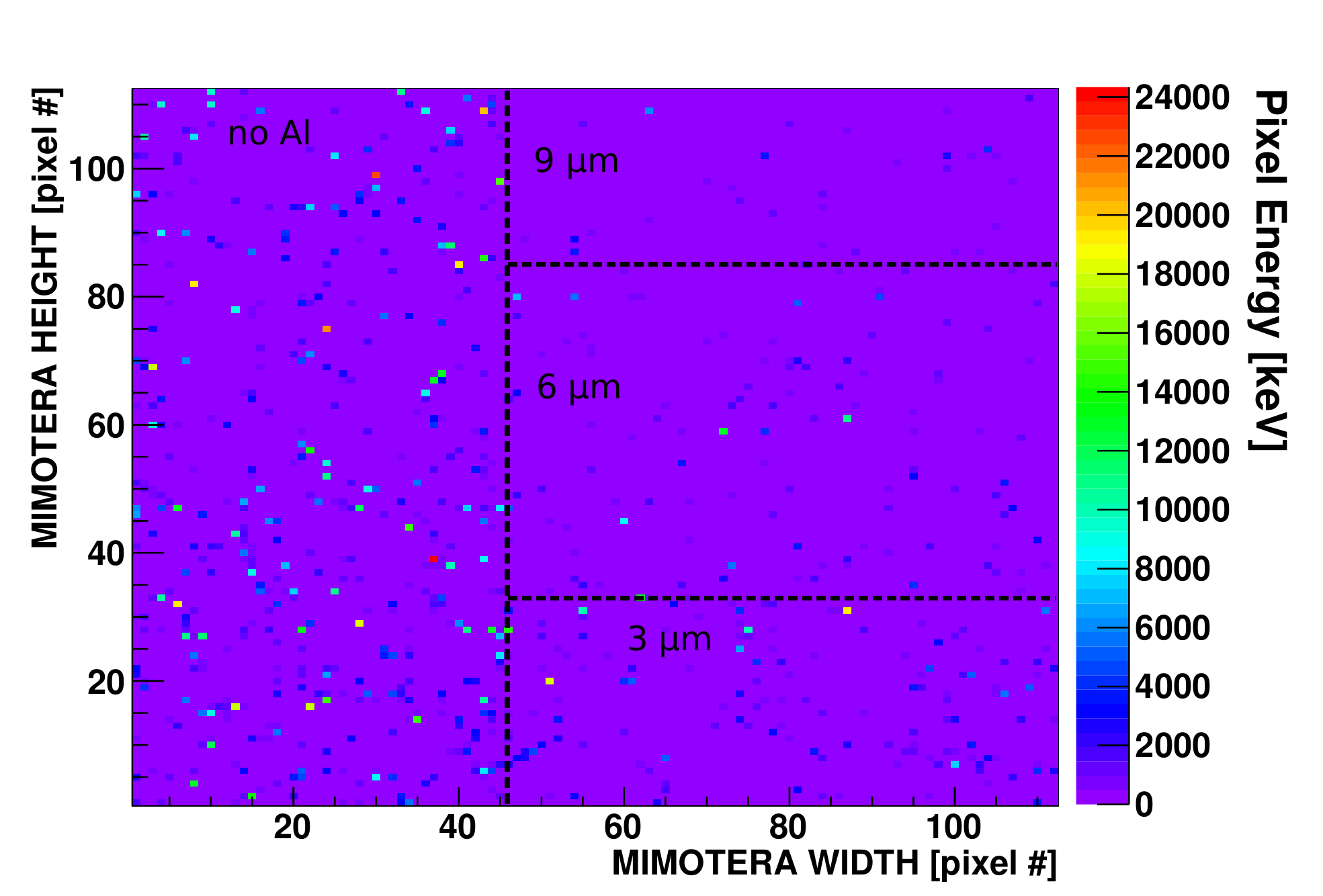}
     \label{fig:01nc}
   }~
   \subfigure[]{
     \includegraphics[width=8cm]{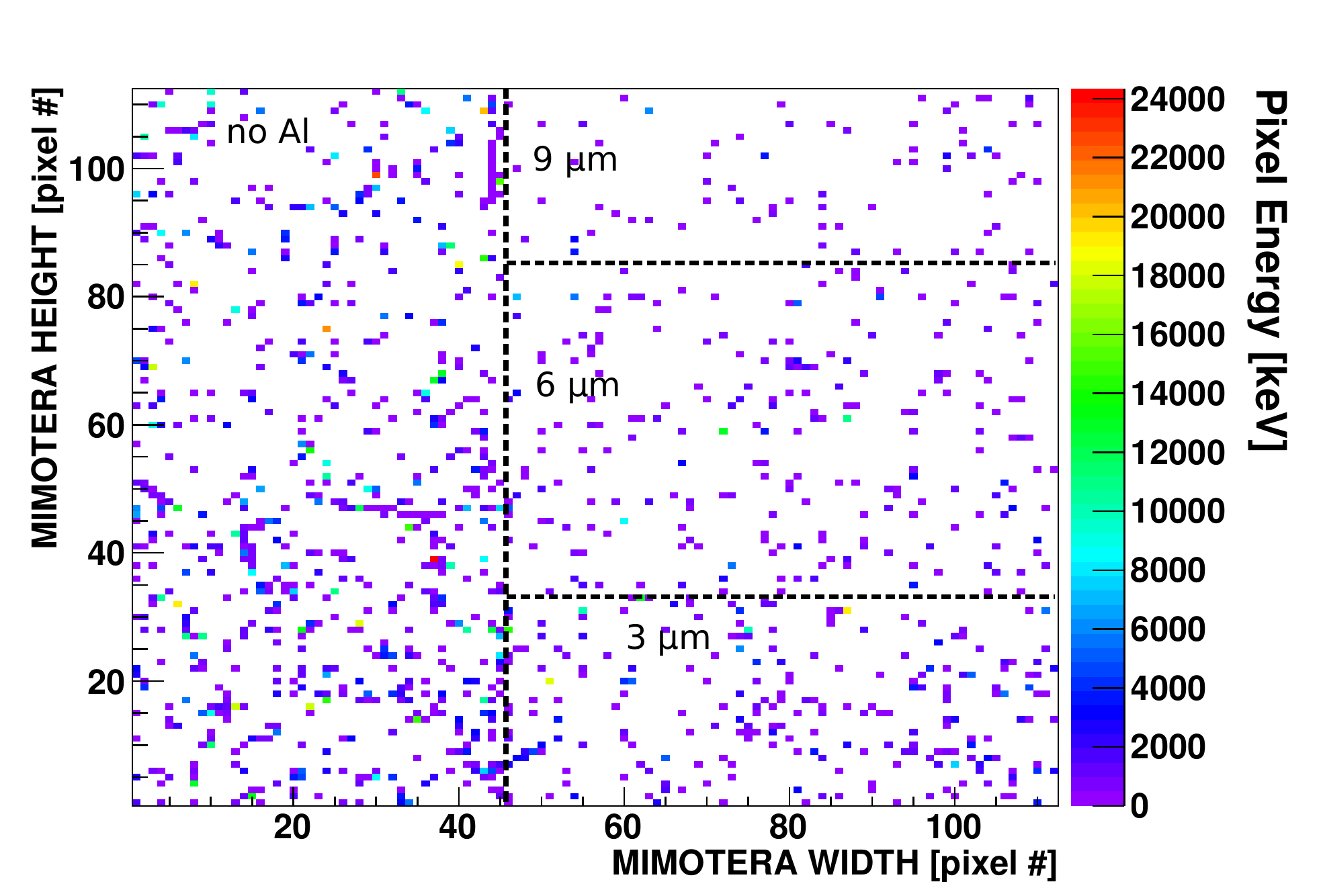}
     \label{fig:01a}
   }
   \centering
   \subfigure[]{
     \includegraphics[width=8cm]{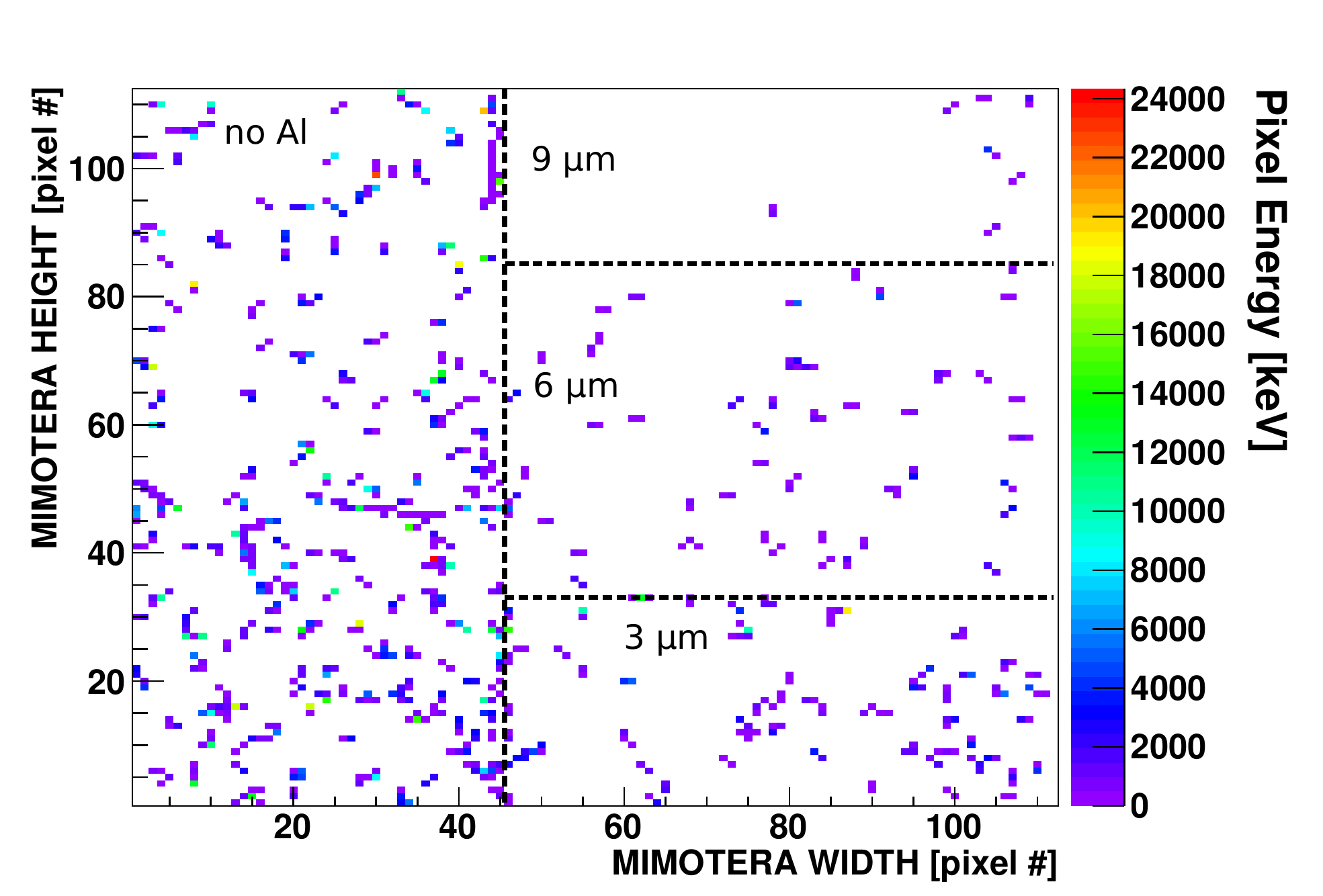}
     \label{fig:01b}
   }
   \caption{Sample of a raw triggered frame (a), after applying the noise cut of 150 keV (b), and with the further exclusion of one-pixel clusters. Around 60\% of the detector was covered with different thicknesses of aluminum foil (3, 6, 9 $\mu$m), as shown in the figures.}
   \label{fig:frames}
\end{figure}

\begin{figure*}
\centering
\begin{minipage}{.5\textwidth}
\centering
\includegraphics[width=8 cm]{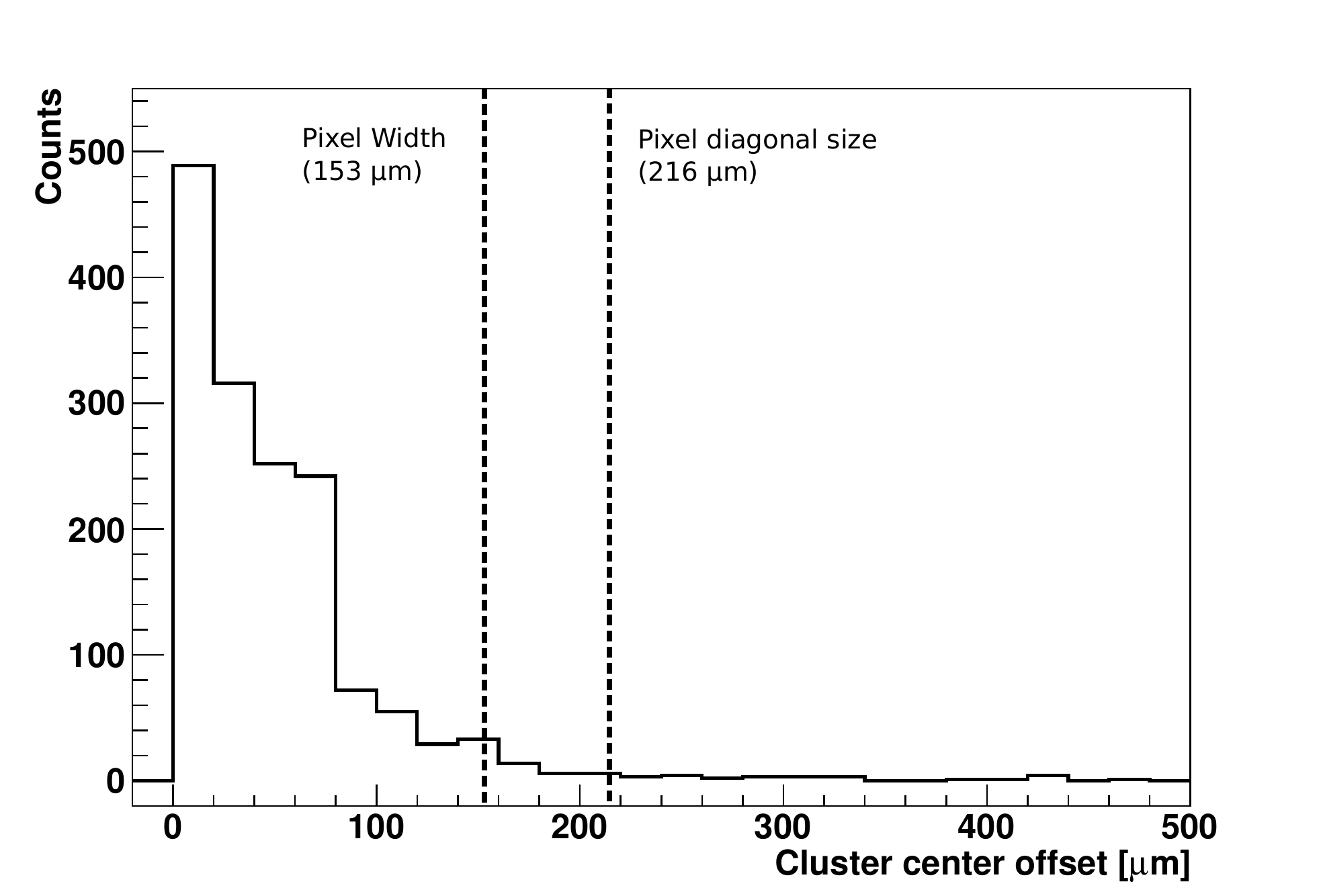}
\caption{Distance offset between the pixel collecting the highest charge in a cluster and the center of gravity of the cluster.}
\label{fig:Offset}
\end{minipage}%
\hspace*{1cm}
\begin{minipage}{.5\textwidth}
  \centering
\includegraphics[width=8cm]{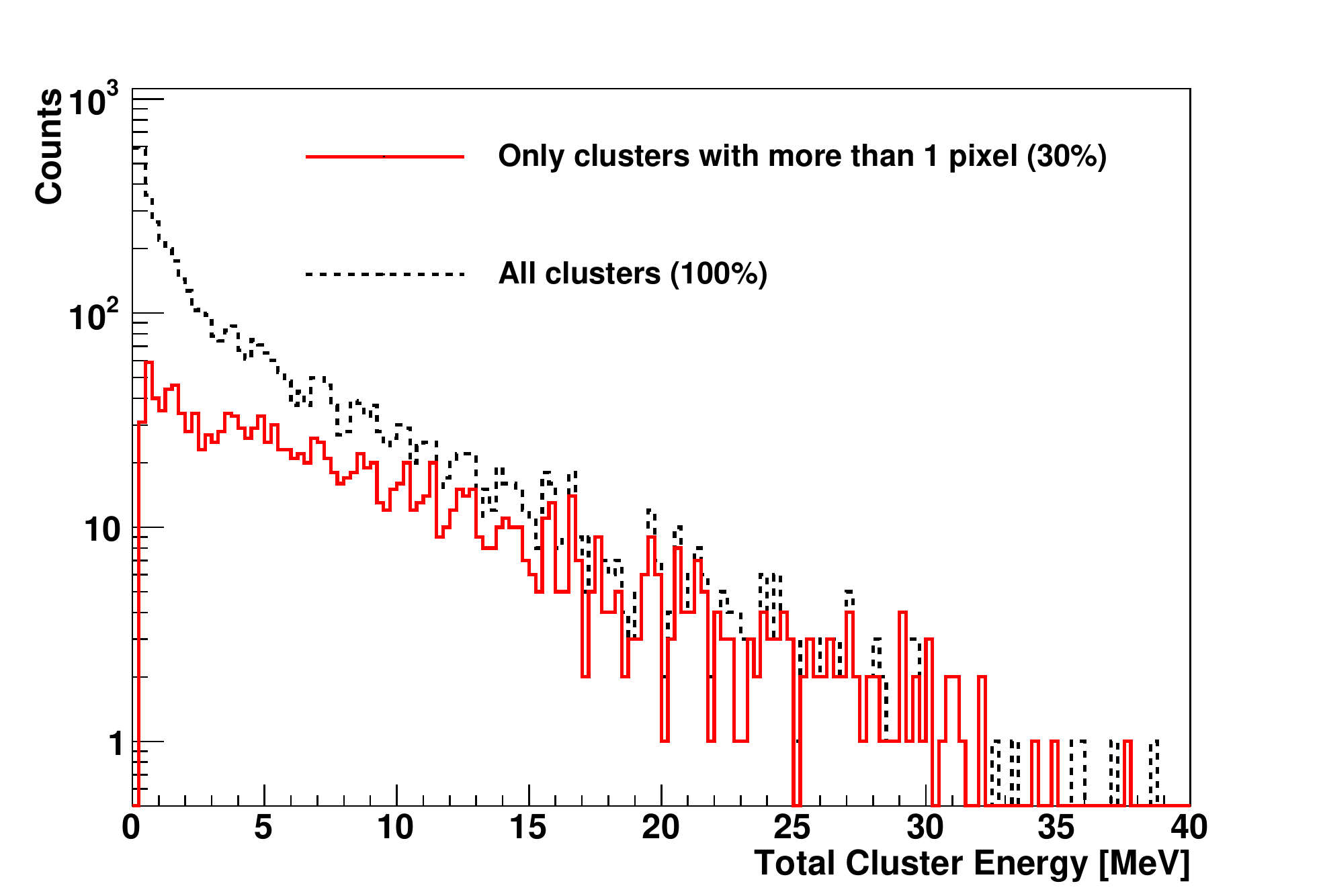}
\caption{Total cluster energy distribution before and after the exclusion of one-pixel clusters (potentially background-affected) from the complete dataset analyzed.}
\label{fig:1pixCutApplied}
\end{minipage}
\end{figure*}

We thus developed a clustering routine tailored to our case. Particles impinging or annihilating in the MIMOTERA were identified by clusters of neighbouring pixels with a signal exceeding 150 keV, i.e. 5 standard deviations of the noise distribution.
Fig. \ref{fig:frames}.a shows a raw frame, while fig. \ref{fig:frames}.b shows the effect of this cut on the same frame. As part of the validation of the clustering algorithm we measured the distance between the center of gravity of each cluster and the pixel collecting the highest charge. The results are shown in fig. \ref{fig:Offset} (for clusters with more than 1 pixel). One can see that 97$\%$ of the clusters have the highest energy pixel coinciding with the geometrical centre. For this reason a seed-driven algorithm using the highest energy pixel of a cluster could possibly be used for future analysis of thin detectors.

\section {Results}

\begin{figure}
   \subfigure[]{
     \includegraphics[width=8cm]{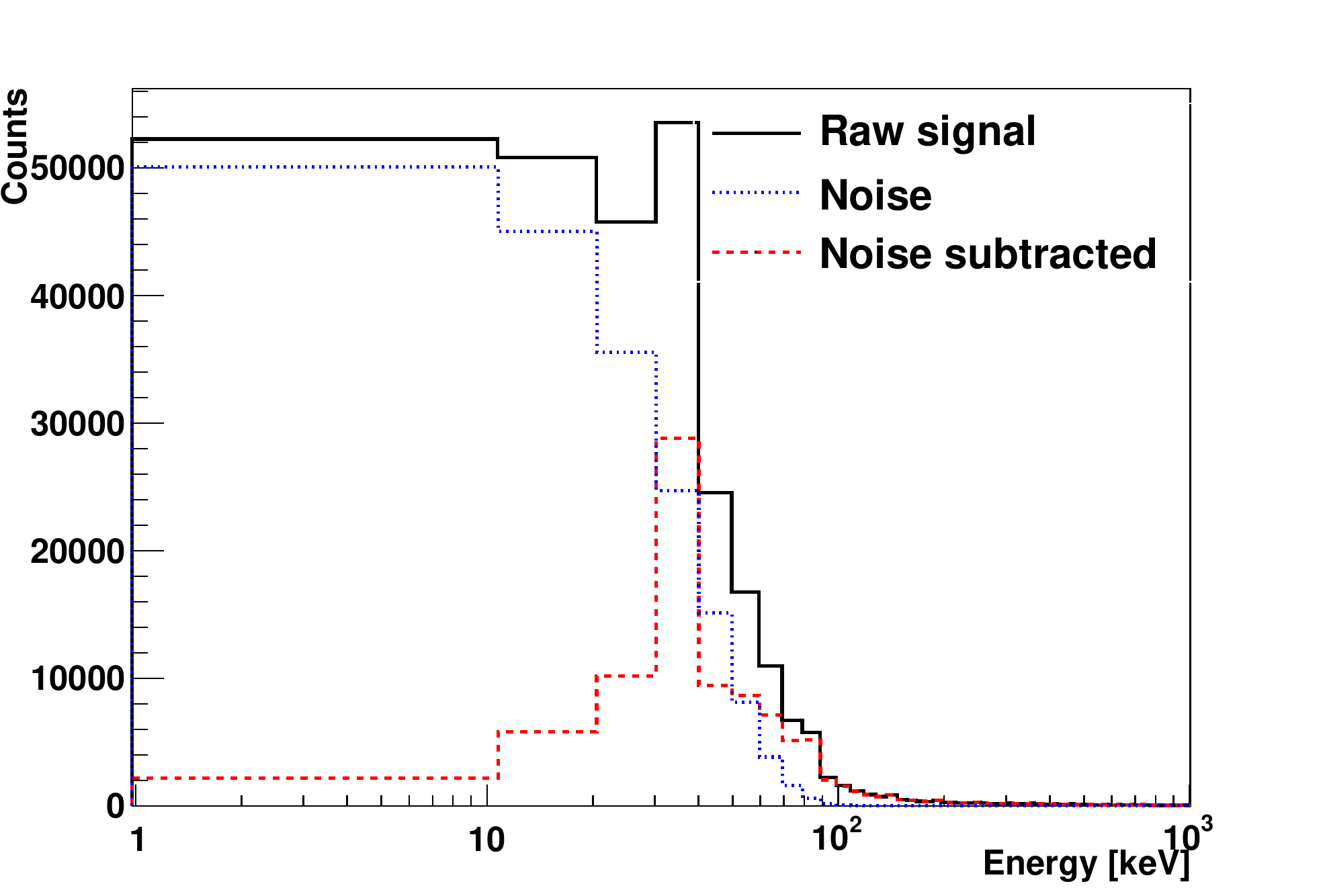}
     \label{fig:pixelenergya}
   }~
   \subfigure[]{
     \includegraphics[width=8cm]{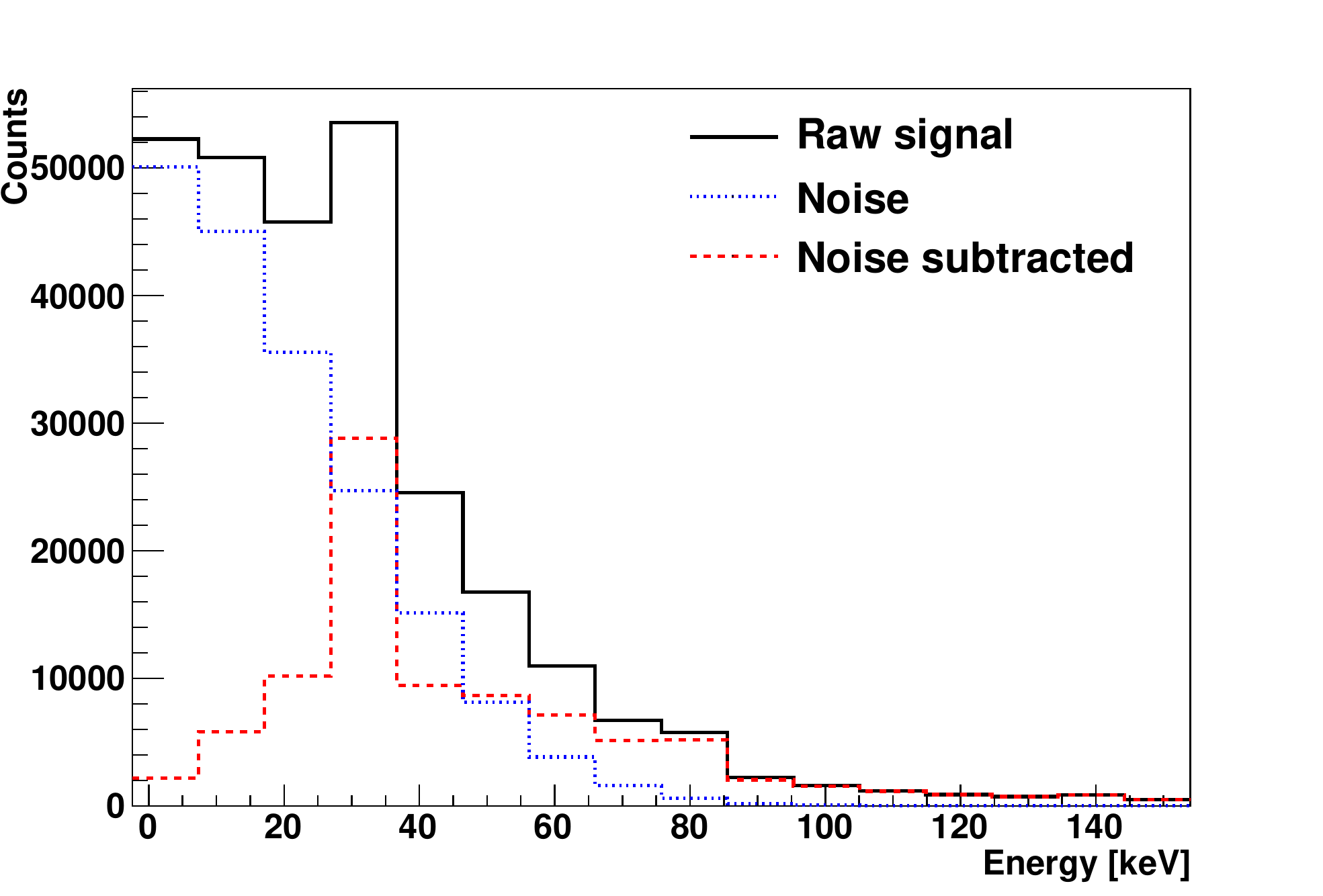}
     \label{fig:pixelenergyb}
   }
   \caption{Distribution of the signal in single pixels after subtraction of the noise fitted with a normal distribution over the whole range of the acquired data (left) and detail of the low energy region (right).}
   \label{fig:pixelenergy}
\end{figure}

The annihilations produce clusters of fired pixels in different shapes and values of deposited energy, up to 20 MeV (see fig.~\ref{fig:pixelenergy}). As many as 20 pixels can be included in a single cluster and some annihilations show one or more tracks coming in from the cluster centre in all directions, in a star shape.


\begin{figure}
\begin{center}
\includegraphics[width=14cm]{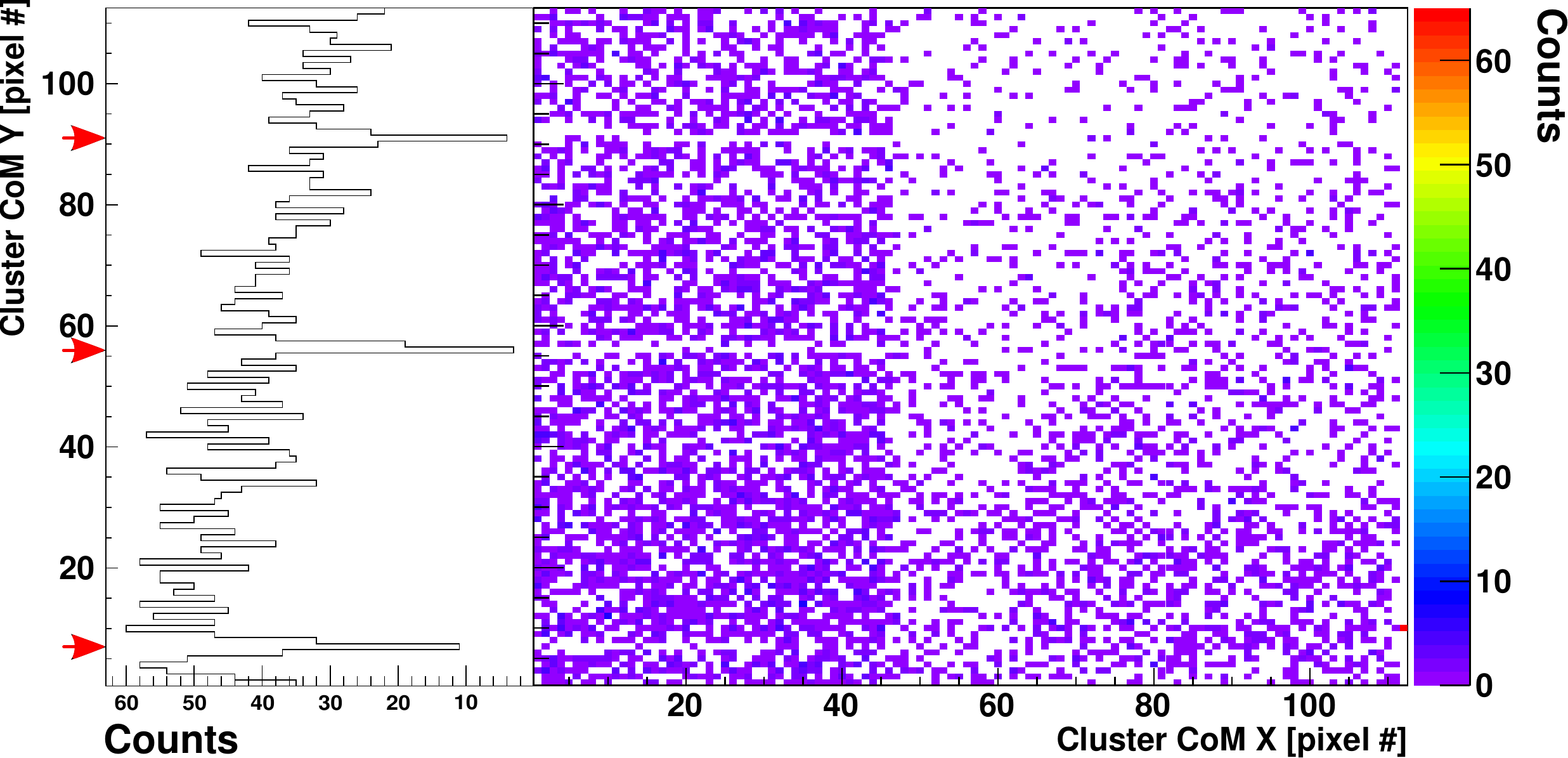}
\caption{Integrated map of cluster center of mass coordinates for the frames used in the analysis, performed for clusters of at least 2 pixels. The red arrows show the location of the supporting wires shadow. Left hand side of the detector was uncovered. The three sections on the right hand side were covered by the 3, 6, 9 $\mu$m aluminum foils, bottom to top. The histogram on the left is the projection of the uncovered part, evidencing the shadows of the wires.}
\label{fig:clustercenterofmassmap}
\end{center}
\end{figure}

\begin{figure}[htp]
\centering
\includegraphics[width=8cm]{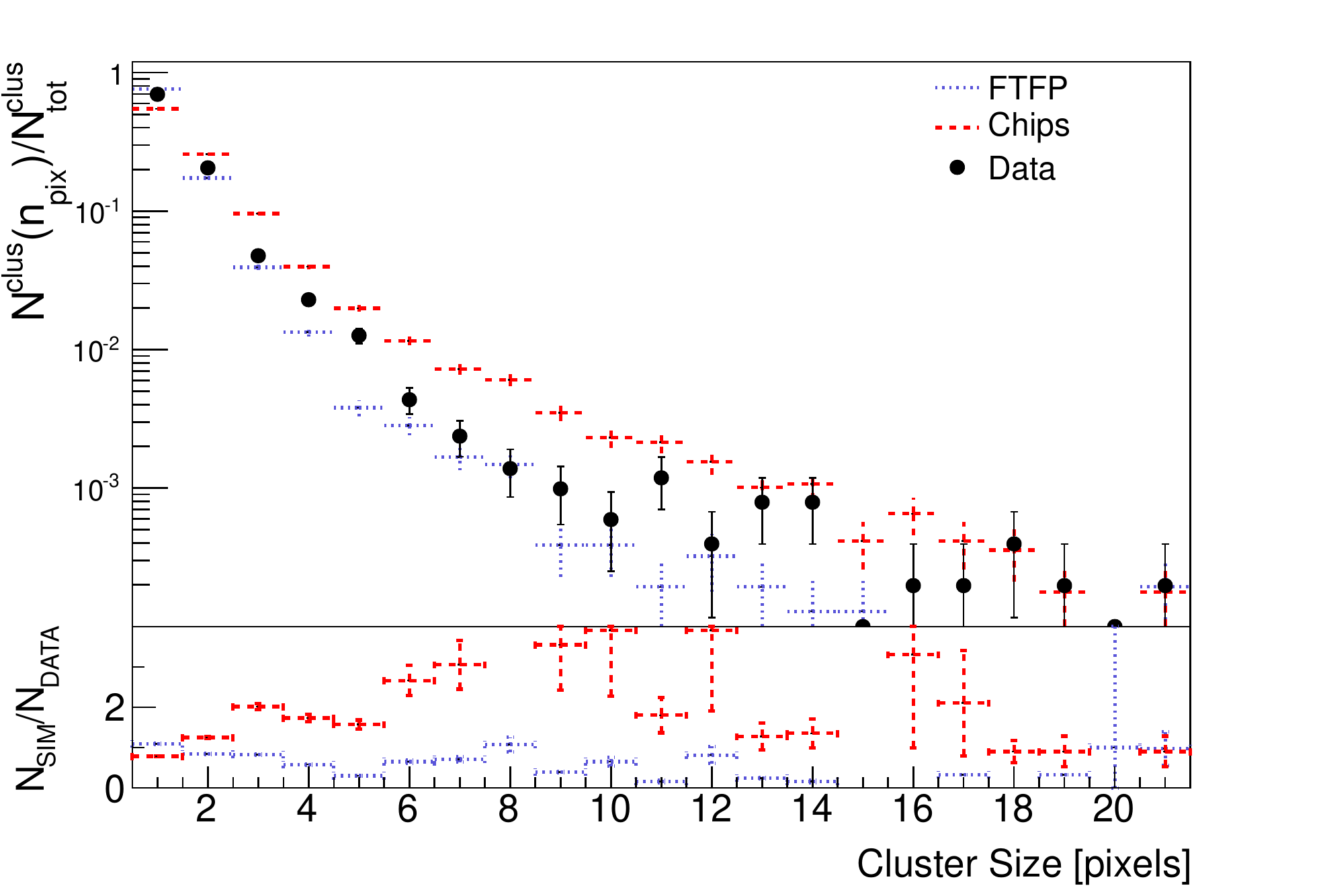}
\caption{Cluster size distribution for data and the two studied simulation models. Most clusters have 1 or 2 pixels, but some clusters consist of as many as 20 pixels.}
\label{fig:NpixClusters}
\end{figure}

\begin{figure*}
\begin{minipage}{.5\textwidth}
\centering
\includegraphics[width=8cm]{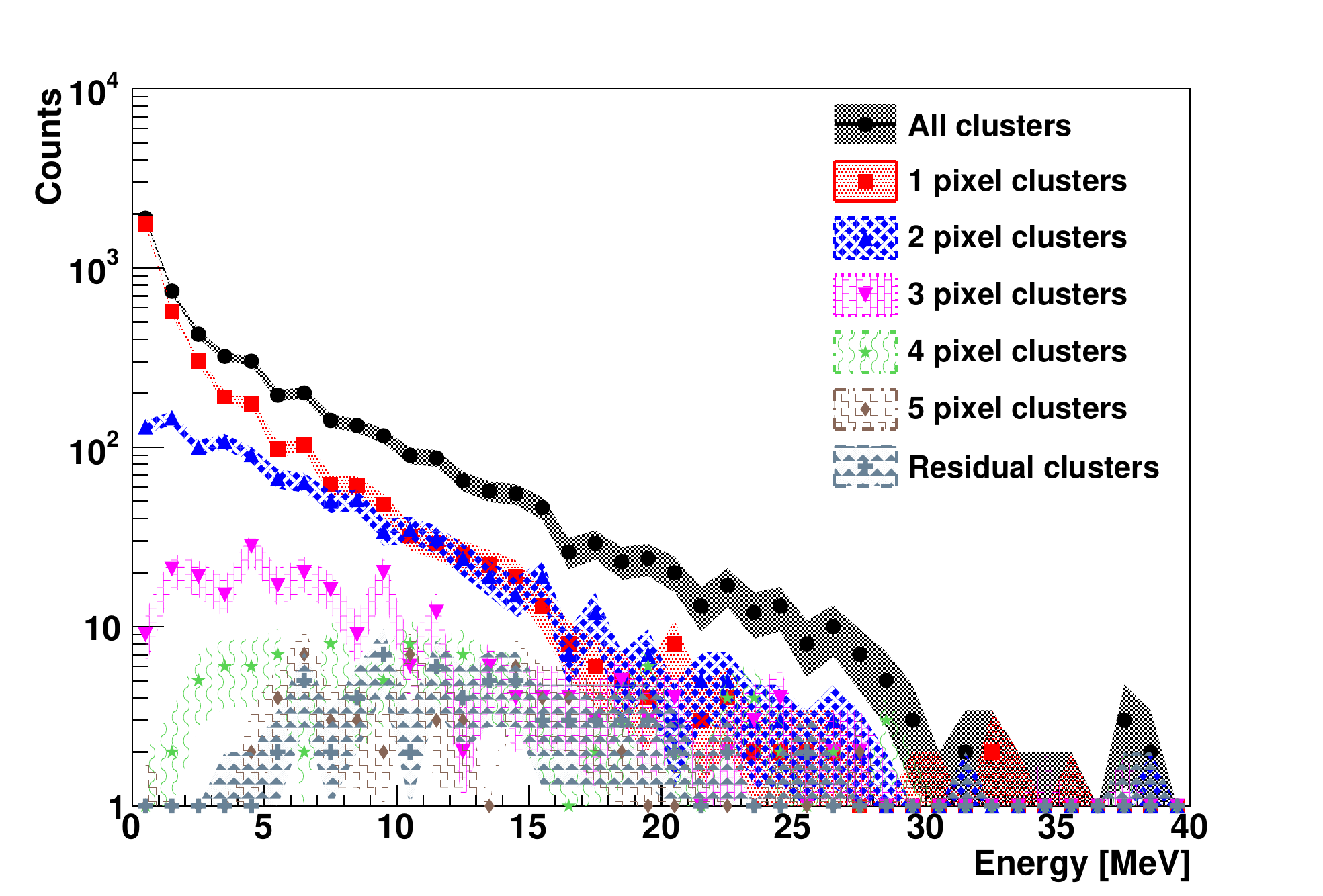}
\caption{Cluster energy spectrum for different cluster sizes showing a large spectrum for all sizes up to 40 MeV. Clusters with few pixels mostly have low energy, while as the clusters get larger they are more evenly distributed. }
\label{fig:EnergyPerSize}
\end{minipage}%
\hspace*{1cm}
\begin{minipage}{.5\textwidth}
\includegraphics[width=8 cm]{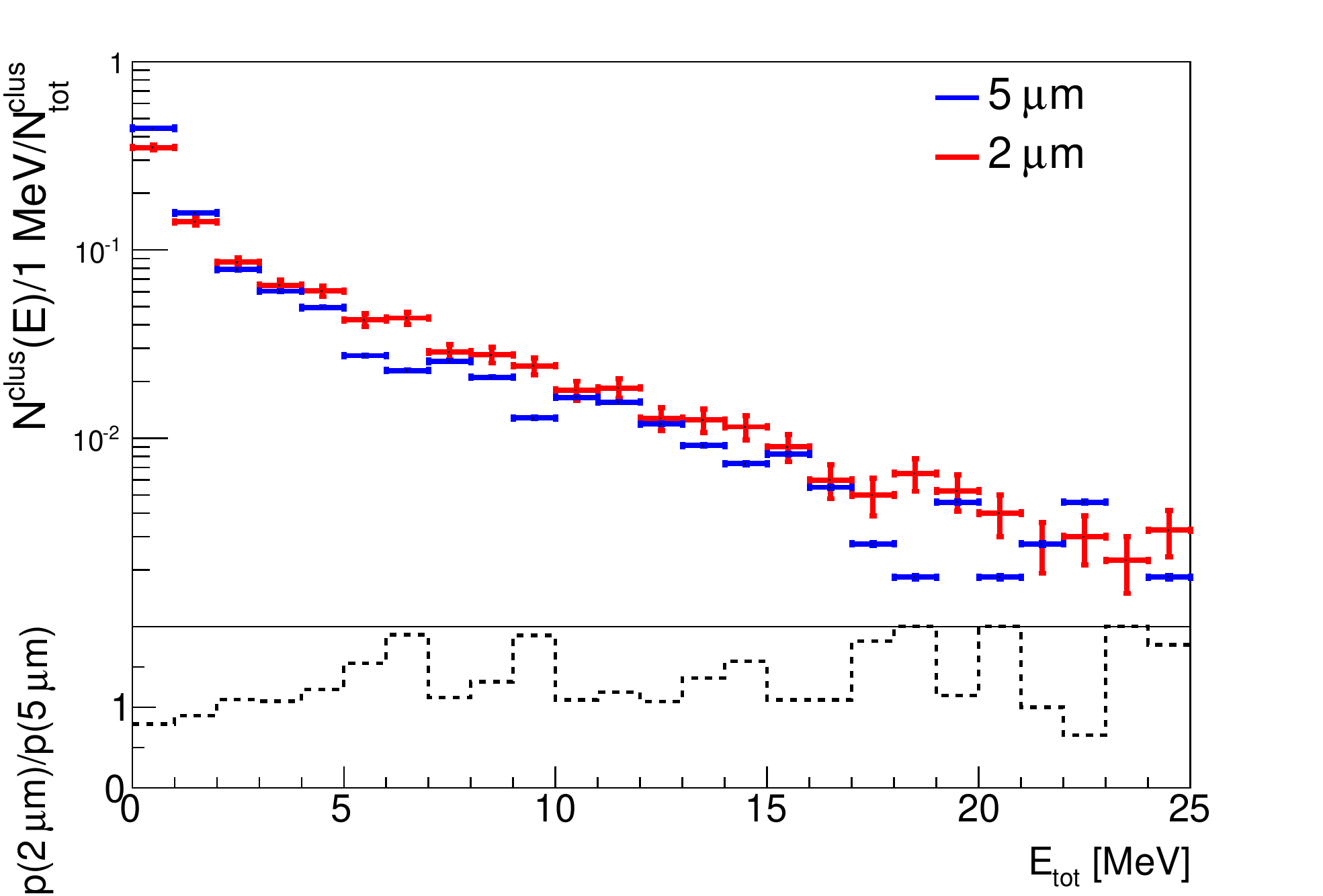}
\caption{Total energy for clusters produced by antiprotons passing through 2 ${\mu}$m and 5 ${\mu}$m thick mobile degrader.}
\label{fig:2_5}
\end{minipage}
\end{figure*}

\begin{figure*}
\begin{minipage}{.5\textwidth}
\includegraphics[width=8cm]{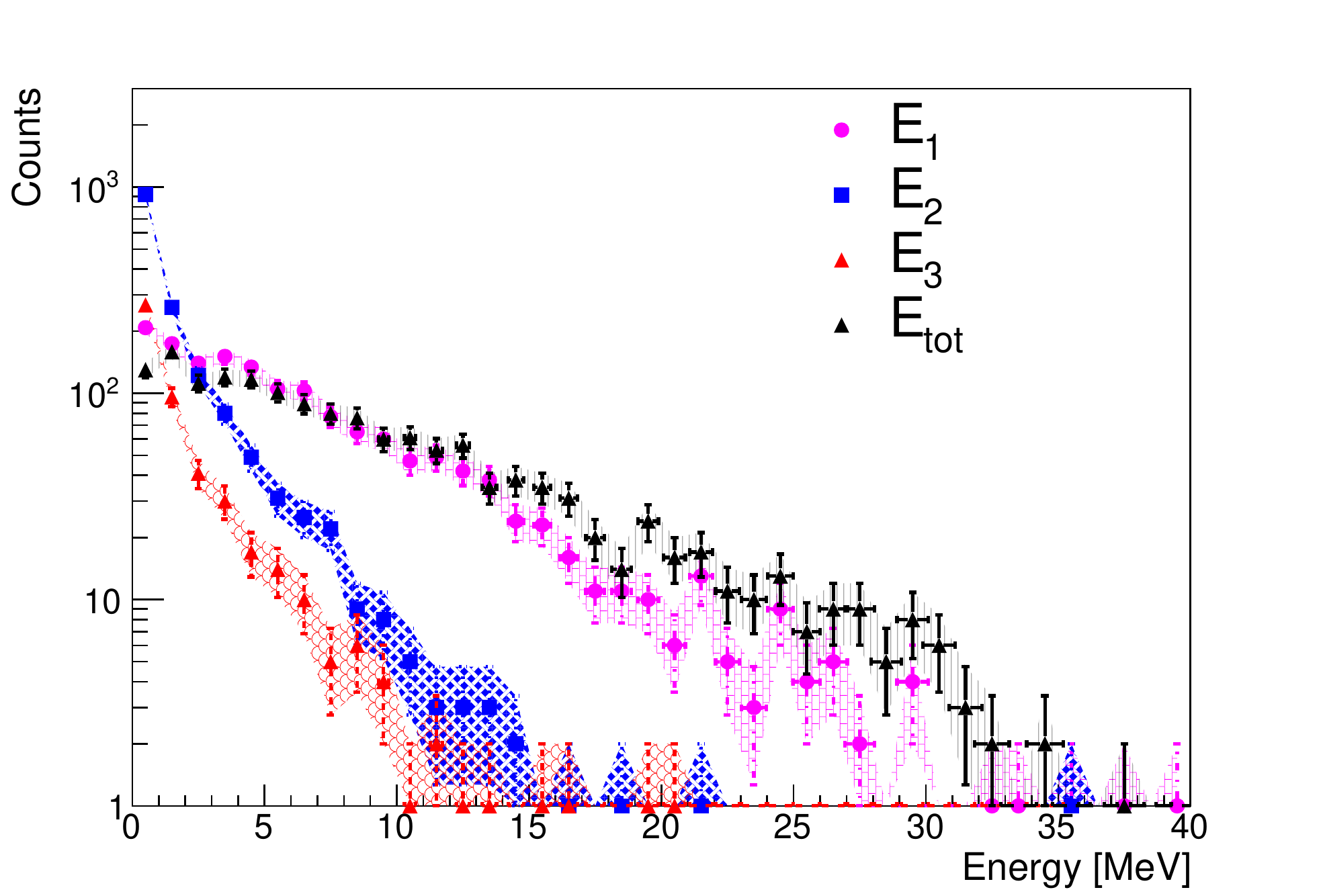}
\caption{Energy distributions for $E_1$, $E_2$, $E_3$, $E_{tot}$ (with $E_1$ energy of the pixel with highest energy in a cluster, $E_2$ energy of the pixel with the second highest energy, $E_3$ the sum of all residual pixels and  $E_{tot}$  total cluster energy) for clusters with more than one pixel.}
\label{fig:AllDistrosgr1}
\end{minipage}%
\hspace*{1cm}
\begin{minipage}{.5\textwidth}
\centering
\includegraphics[width=8cm]{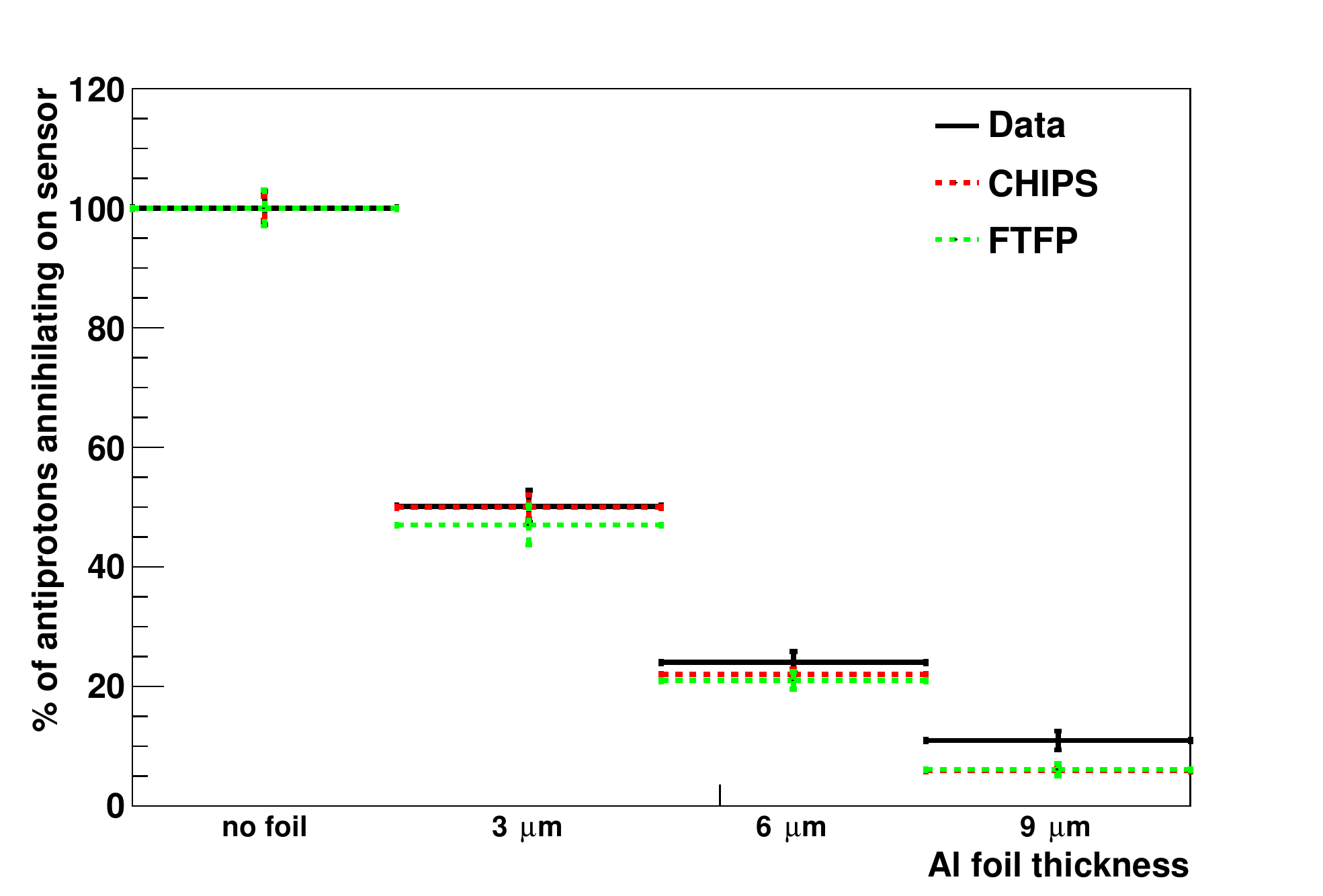}
\caption{Fraction of annihilations on the sensor parts covered with different thicknesses of Al foil. This analysis is performed only with clusters consisting of more than 3 pixels.}
\label{fig:AnnihilationsOnSensor}
\end{minipage}
\end{figure*}

\subsection{Data selection}

The efficiency of the clustering algorithm strongly depends on the probability of having two or more overlapping clusters. For this reason, a veto was applied on frames with too high pixel or cluster occupancy. Occupancy was varying throughout the data taking because of different configurations of the focusing magnet. Only frames with a pixel occupancy $<$ 10 $\%$ and less than 150 clusters per frame were accepted, resulting in $\sim$ 25 \% of the frames being included in the analysis.

\subsection{Background sources}

\label{ss:background}

Two possible background sources were identified as potentially affecting the acquired data. 
In AE$\mathrm{\bar{g}}$IS, heavy ions and protons produced from annihilations in the central region of the apparatus are one of the possible background sources. They are expected to arrive quasi-normally on the sensor at an angle of $\sim$ 0.1 rad with an estimated 1$\%$ probability to produce clusters with a size exceeding 1 pixel. It is worth remarking that (see table \ref{tab:wiredfractions}) about one third of the total clusters observed were composed by more than one pixel.

\begin{table}
\centering
\begin{tabular}{|L|L|L|}\hline
Min. Clus. Size (pix.) & N. of clusters & Cluster ratio in shadowed area ($\%$)\\\hline
1&11 537&16.0$\pm$3.4\\
2&4 401&7.4$\pm$3.6\\
3&1 911&9.3$\pm$6.4\\
4&1 056&8.2$\pm$7.7\\
\hline
\end{tabular}
\caption{Fraction of clusters centers of mass in rows shadowed by wires with respect to clusters in neighboring rows, for the region not covered by Al foils.}
\label{tab:wiredfractions}
\end{table}


Two pixels clusters can be generated by a background source only if the source is not quasi-normal to the detector itself. The shadow of the wires used to support the Al foils partially masking the MIMOTERA was used to estimate the fraction of particles impinging on the detector along directions other than normal, most likely due to annihilations in the apparatus. Fig. \ref{fig:clustercenterofmassmap} shows the map of centre of mass (CoM) for all clusters with at least 2 pixels. The shadows left by the wires (300 $\mu$m gauge) are clearly visible. The wire's geometrical shadow on the sensor can be calculated to cover an angle of $3.5^\circ$. This angle is quoted with respect to the average direction of the incoming antiprotons (see sec. \ref{ss:facility}). Table \ref{tab:wiredfractions} shows the ratios between the number of clusters in shadowed and unshadowed rows for different cluster sizes. While the contamination for single pixel clusters is at the $16.0\%$ level, it drops to the $7\%$ level for larger clusters.


These numbers set a limit for the purity of the sample by particles travelling with high divergence from the antiproton flux. This constrast ratio also represents the contamination limit for particles different from antiprotons. Any further cut on the number of pixels doesn't introduce any significative improvement, while reducing the statistics.
Fig. \ref{fig:frames}.c shows a sample frame after the cut on single pixel clusters, and fig. \ref{fig:1pixCutApplied} shows the effect of this cut on the cluster energy distribution.

For the reasons exposed above, a more detailed analysis on the energy and size of the clusters and comparison with simulations will only be shown for the sample with highest purity, i.e. the one composed by clusters with at least two pixels.

\subsection{Cluster characteristics}

Fig. \ref{fig:NpixClusters} shows the distribution of cluster sizes for both data and the two simulation models. Although we find clusters as big as 20 pixels, $\sim 2/3$ of the events are formed by one pixel  and $\sim 1/3$ of two or more pixels, indicating a prevalence of localized energy deposits.

The total cluster energy spectrum is seen in fig. \ref{fig:EnergyPerSize}, showing cluster energies as large as 40 MeV. This figure also shows the energy distribution of clusters of different sizes, and one can see that the energies of a given cluster size are distributed over the entire energy range. Small clusters are most often produced at low energies, with a sloped distribution decreasing towards higher energies. As the size of the clusters increases, the slope of the energy distribution flattens out and the minimum energy is shifted upwards, starting above $\sim1$ MeV for clusters with 4 pixels or more.

Since data were taken with two different degrader configurations, the datasets were studied in order to verify whether there was enough statistically significant difference to justify a separate analysis.
We separated the events collected with 2 and 5 ${\mu}$m degrader, and the corresponding energy spectrum can be seen in fig. \ref{fig:2_5}. 
The overall distribution of the energy of the annihilation clusters is very similar for both degraders. Small statistically significant differences can be observed for only a few bins, probably related to the differences in the penetration depth (and hence the deposited kinetic energy) in silicon for the two degraders configuration. However, given the non systematicity of the difference, we decided to consider the two datasets together, thus improving significantly the statistics.

\begin{figure}
\begin{center}
\includegraphics[width=8cm]{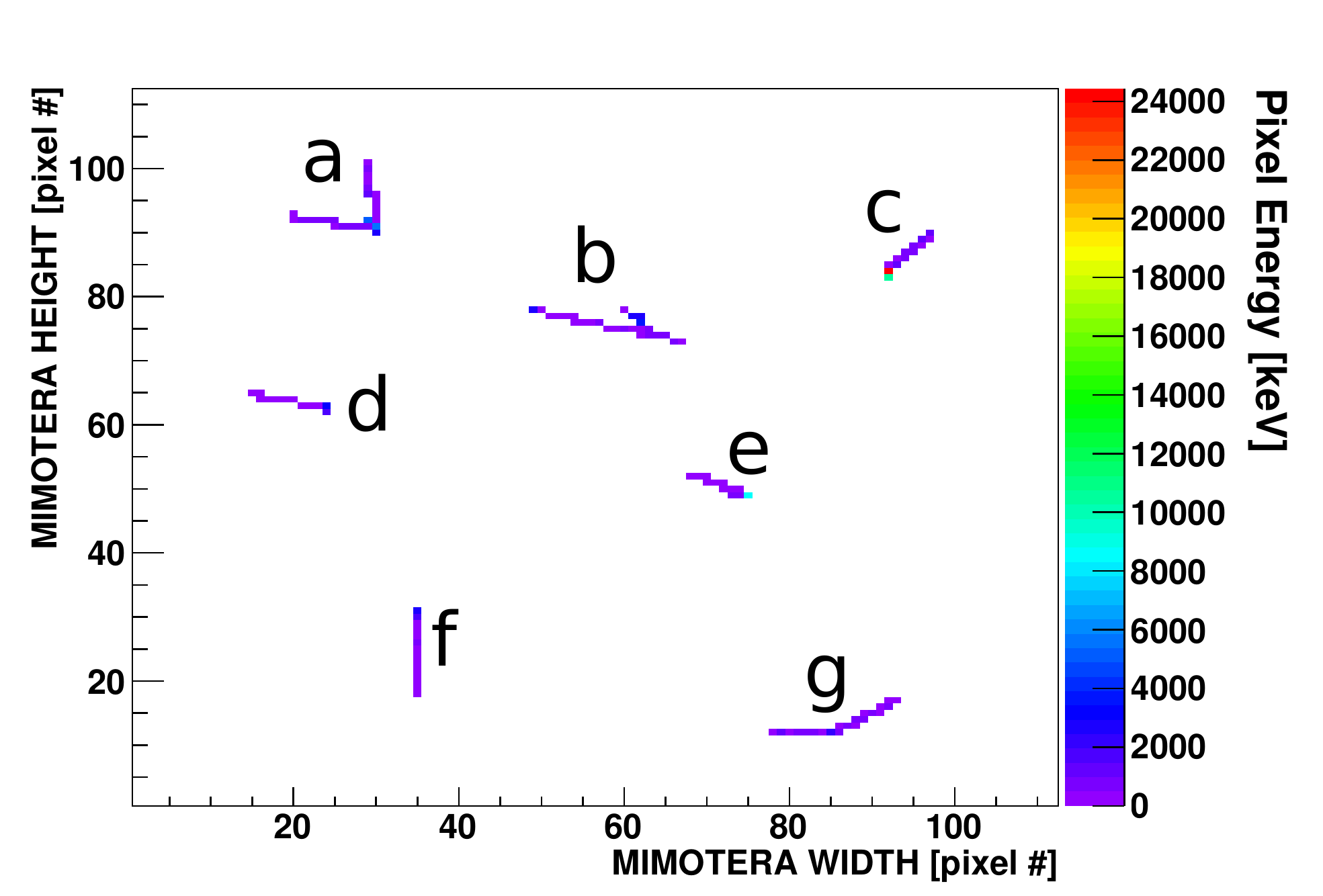}
\caption{Sample of in-plane tracks observed with the MIMOTERA detector. A description of the tracks is provided in table \protect\ref{tab:id}.}
\label{fig:tracks}
\end{center}
\end{figure}

Fig. \ref{fig:AllDistrosgr1} shows how the energy is distributed among the pixels composing the clusters. $E_1$ is the energy distribution of the pixel with the highest energy in the cluster, $E_2$ the pixel with the second highest energy, $E_3$ the residual energy and $E_{tot}$ the total cluster energy. These distributions are shown after cutting on 1 pixel clusters. For all clusters, most of the energy is concentrated in one single pixel. 

The additional Al foils covering the detector were used to study the energy loss of antiprotons in silicon. To be sure to study antiprotons which annihilated inside the silicon, we counted  clusters with more than 3 pixels only. This cut removes not only single pixel clusters from particles annihilating in the foils or elsewhere in the apparatus but also excludes the rare case where a secondary particle passes at the intersection of 2 pixels. Since we have shown that data taken with the two degraders were compatible (fig. \ref{fig:2_5}), the results which include the additional Al foil are shown in fig. \ref{fig:AnnihilationsOnSensor} for both types of degraders $-$ 2 and 5 ${\mu}$m. About  $\sim10\%$  of the antiprotons were able to pass through the 9 ${\mu}$m aluminum foil. The results for the different aluminum thicknesses are in agreement with both simulation models, showing that the models provide a good description of the stopping power of antiprotons in matter.

\subsection{Tracks recognition}
\label{tracks}
Measuring track lengths and ${dE}/{dx}$ proved to be a useful method to identify some of the annihilation products travelling in the silicon detector. 
Given the small thickness of the MIMOTERA active region, products traveling in  the detector  plane were scarce. However, we were still able to distinguish 21 clusters having one, two or three ion tracks.

To identify the annihilation products we calculated the ranges and $dE/dx$ for the most important ion species produced in the annihilation process \cite{Markiel1988}. As mentioned in sec. \ref{Montecarlo}, fig. \ref{fig:EnergyDeposition} shows the deposited energies and fig. \ref{fig:StoppingRange} shows the corresponding ranges.  For heavy ion species with low energies, where the range is $<$14 ${\mu}$m,  the total particle kinetic energy is expected to be deposited in the detector. 

Fig. \ref{fig:tracks} shows examples of typical clusters with tracks, and table \ref{tab:id} lists the properties of all tracks found.   From the deposited energy most of the  tracks can be identified as protons, while one track probably originates from a heavier ion.

\begin{table*}[htp]
\centering
\begin{tabular}{|S|S|L|L|V|}\hline
N. of prongs & Seed energy (MeV) & Prongs length ($\mu$m) & Prongs $dE/dx$ (keV/$\mu$m) & Identification\\\hline
1&none&1630&1.18&Proton $>$ 100 MeV\\
1&9.9&2950&2.11&Proton $\simeq$ 50 MeV\\
$^{(d)}$1&4.9&1650&1.63&Proton $\simeq$ 70 MeV\\
$^{(c)}$1&34.8&1080&6.66&Proton $\simeq$ 10 MeV\\
1&2.6&1840&2.73&Proton $\simeq$ 40 MeV\\
1&17.6&2170&7.06&Proton $\simeq$ 10 MeV\\
$^{(f)}$1&4.4&1840&2.44&Proton $\simeq$ 40 MeV\\
1&8.8&2300&1.34&Proton $\simeq$ 100 MeV\\
1&none&1740&2.16&Proton $\simeq$ 50 MeV\\
1&none&7220&1.12&Proton $>$ 100 MeV\\
1&12.1&2170&2.8&Proton $\simeq$ 30 MeV\\
1&8.6&1730&2.1&Proton $\simeq$ 50 MeV\\
1&6.9&1780&1.2&Proton $>$ 100 MeV\\
1&2.3&2380&2.8&Proton $\simeq$ 40 MeV\\
1&11.1&2190&3.4&Proton $\simeq$ 30 MeV\\
1&none&2900&3.1&Proton $\simeq$ 30 MeV\\
$^{(e)}$1&none&1220&3.0&Proton $\simeq$ 30 MeV\\
$^{(g)}$2&2.2&1100, 1500& 3.9, 3.27& Protons $\simeq$ 30 MeV\\
2&11.1&340, 2080& 0.7, 1.2 & Protons $>$ 100 MeV\\
$^{(a)}$2&13.5&1510, 1620& 4.0, 2.4 & Protons $\simeq$ 20, 50 MeV\\
$^{(b)}$3&none&2200, 900, 750& 2.4, 4.1, 15 & Prot. (50, 20 MeV) + Heavy Ion\\
\hline
\end{tabular}
\caption{Clusters which are identified as having one or more tracks. Clusters marked with a letter are shown in fig. \protect\ref{fig:tracks}. Seeds are here defined as pixels located at one end of the track(s) with pixel energy in excess of 1 MeV.}
\label{tab:id}
\end{table*}

\subsection{Comparison with Monte Carlo simulations}

The Monte Carlo samples were generated separately for CHIPS and FTFP and consist of 3 million events each. The entire flight path of the antiprotons was simulated, starting with the 5.3 MeV antiprotons from the AD, including all of the AE$\mathrm{\bar{g}}$IS apparatus (full geometry and 5 T magnetic field), ending with the annihilations on the silicon detector. In the nominal case, with 225 ${\mu}$m total degrader thickness, only $\sim$ 25 000 antiprotons of the original 3 million annihilated on the detector according to the simulations. For 229 ${\mu}$m thickness this number decreased to $\sim$ 20 000.

\begin{figure*}
\centering
\begin{minipage}{.5\textwidth}
  \centering
\includegraphics[width=8cm]{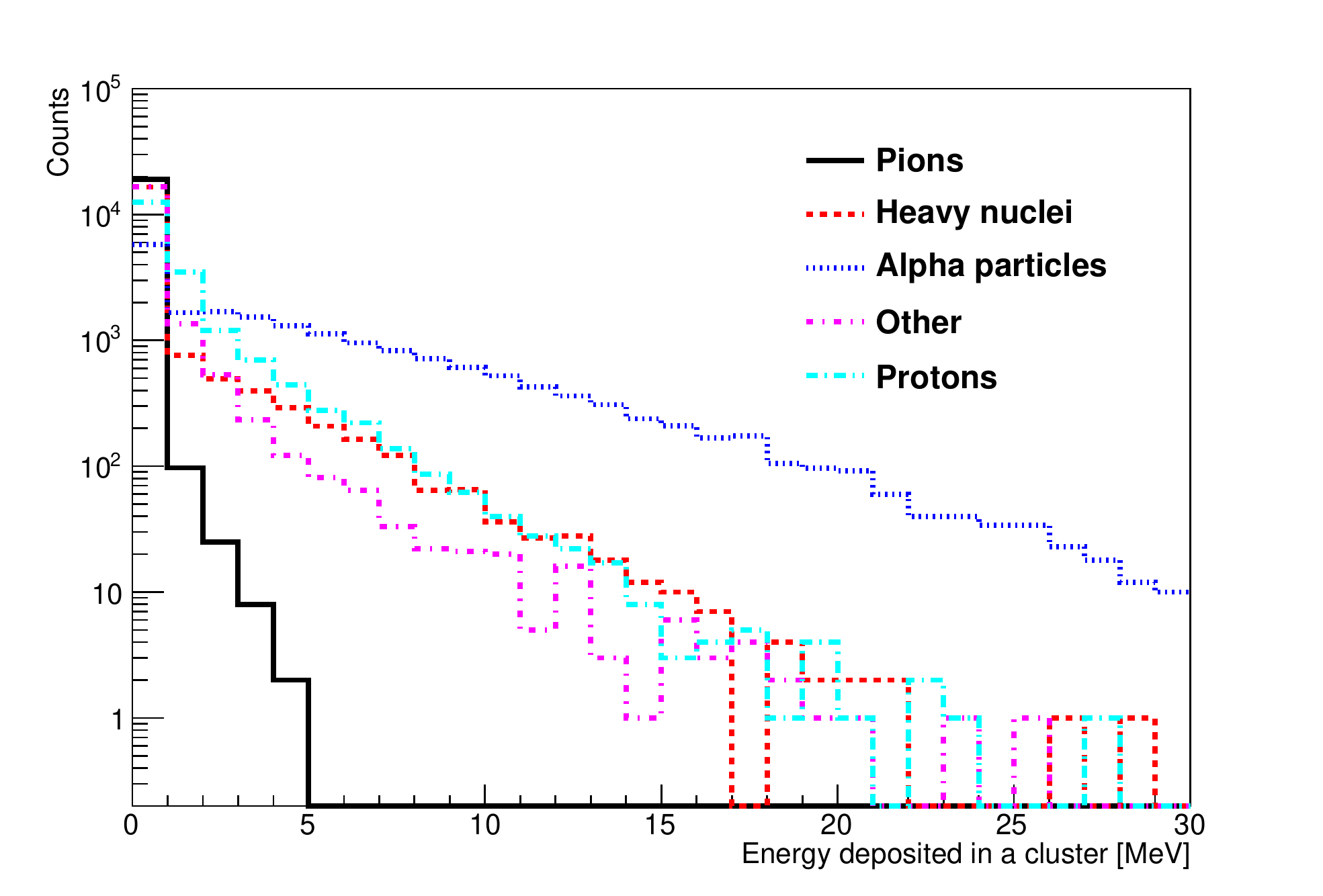} 
\caption{Fraction of cluster energy for different annihilation products, as simulated with CHIPS.}
\label{fig:StackPlotsChips}
\end{minipage}%
\hspace*{1cm}
\begin{minipage}{.5\textwidth}
  \centering
\includegraphics[width=8cm]{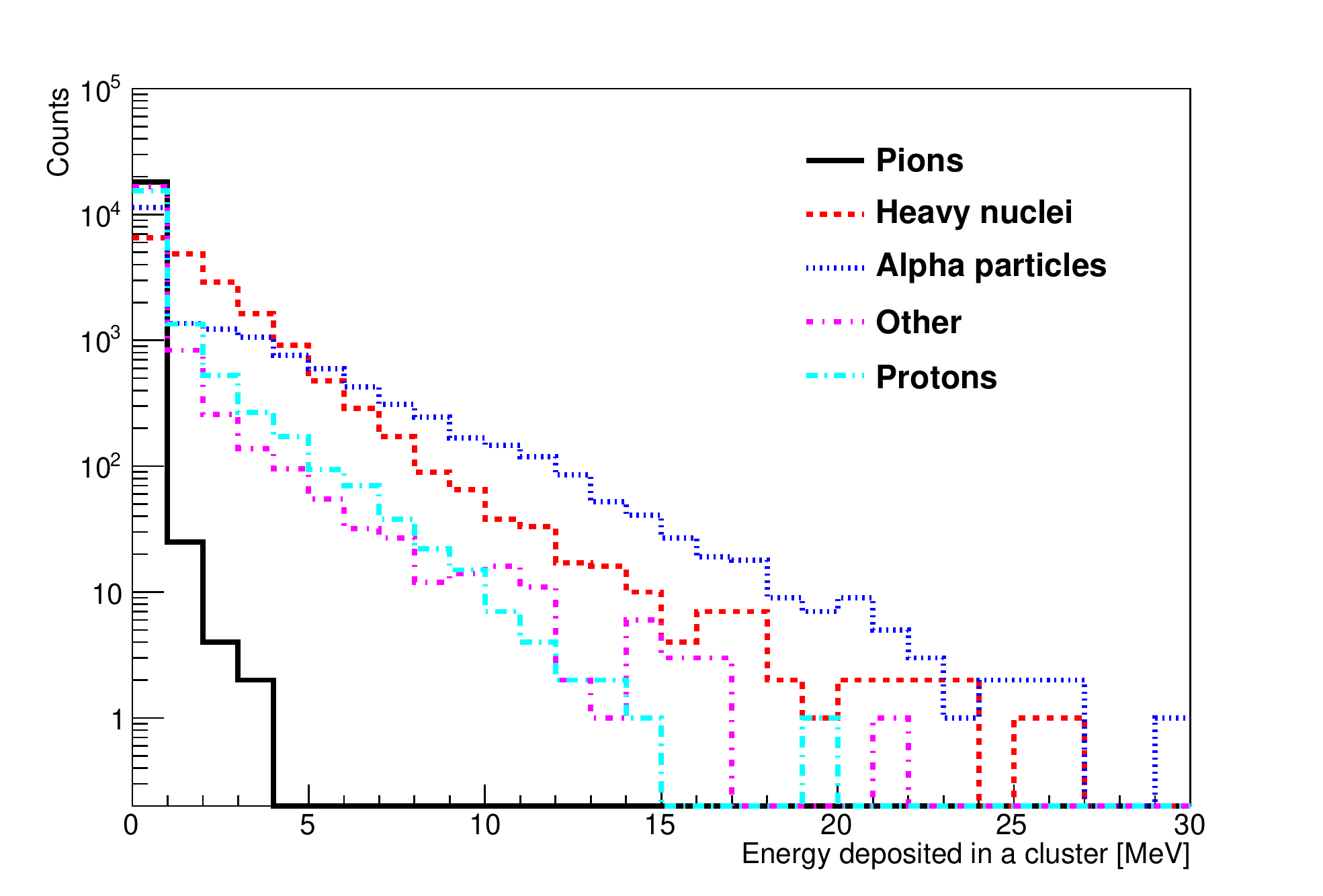} 
\caption{Fraction of cluster energy for different annihilation products, as simulated with FTFP.}
\label{fig:StackPlotsFTFP}
\end{minipage}
\end{figure*}

Fig. \ref{fig:StackPlotsChips} and \ref{fig:StackPlotsFTFP} show the total energy distribution and the particle composition of clusters for the two simulation models. For CHIPS one expects higher cluster energies and a broader distribution containing more alpha particles and protons and less heavy nuclei than for FTFP.

The signal in single pixels was obtained from the ionizing energy deposited by particles in the geometrical volume covered by the pixel cell.

\begin{figure*}[htp]
\centering
\begin{minipage}{.5\textwidth}
  \centering
\includegraphics[width=8cm]{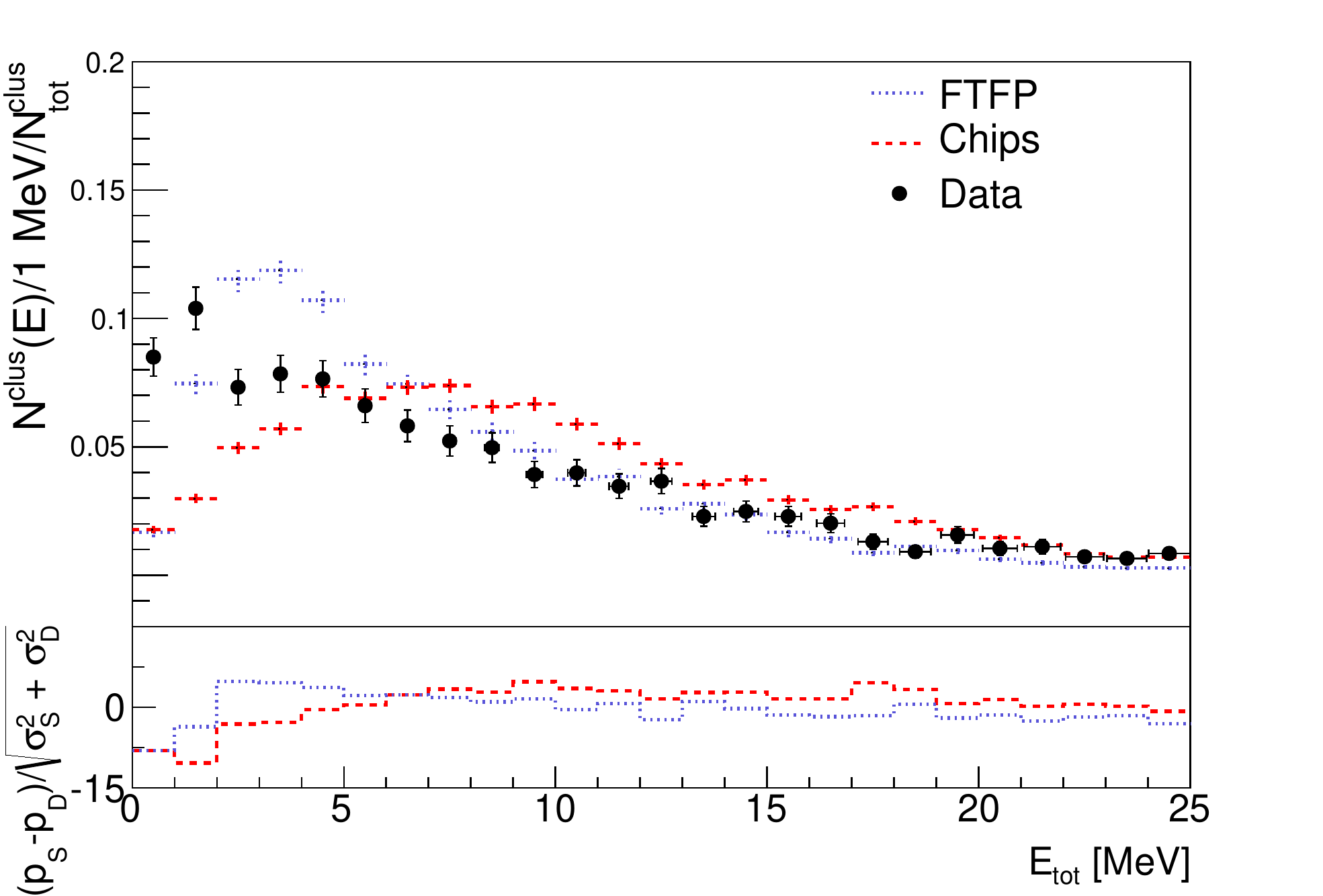}
\caption{Distribution of the total cluster energy, excluding the one-pixel clusters.}
\label{fig:Etotalgr1}
\end{minipage}%
\hspace*{1cm}
\begin{minipage}{.5\textwidth}
  \centering
\includegraphics[width=8cm]{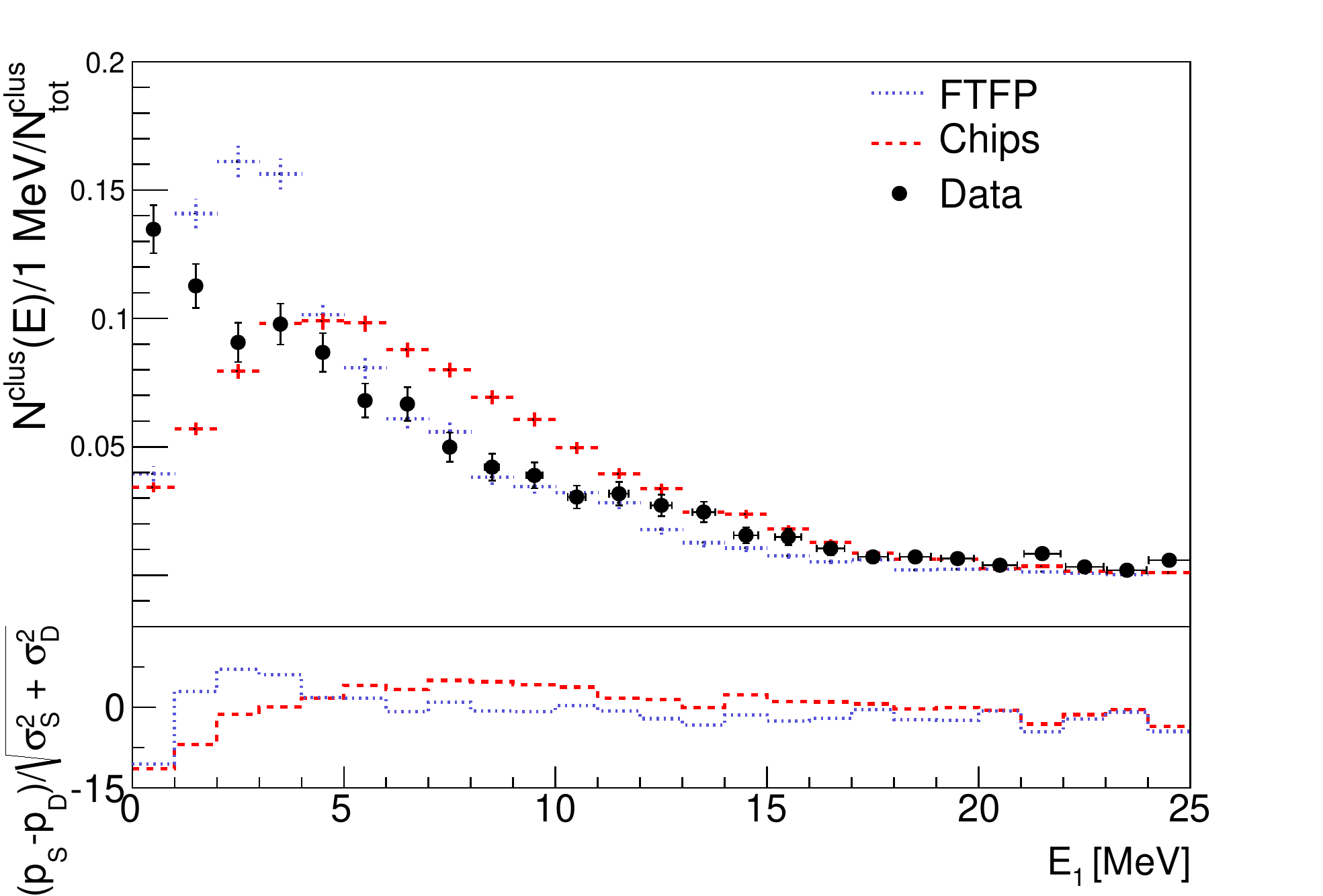}
\caption{Distribution of the energy deposited only in the pixel collecting the highest charge for each cluster, excluding the one-pixel clusters.}
\label{fig:E1gr1}
\end{minipage}
\centering
\begin{minipage}{.5\textwidth}
  \centering
  \includegraphics[width=8cm]{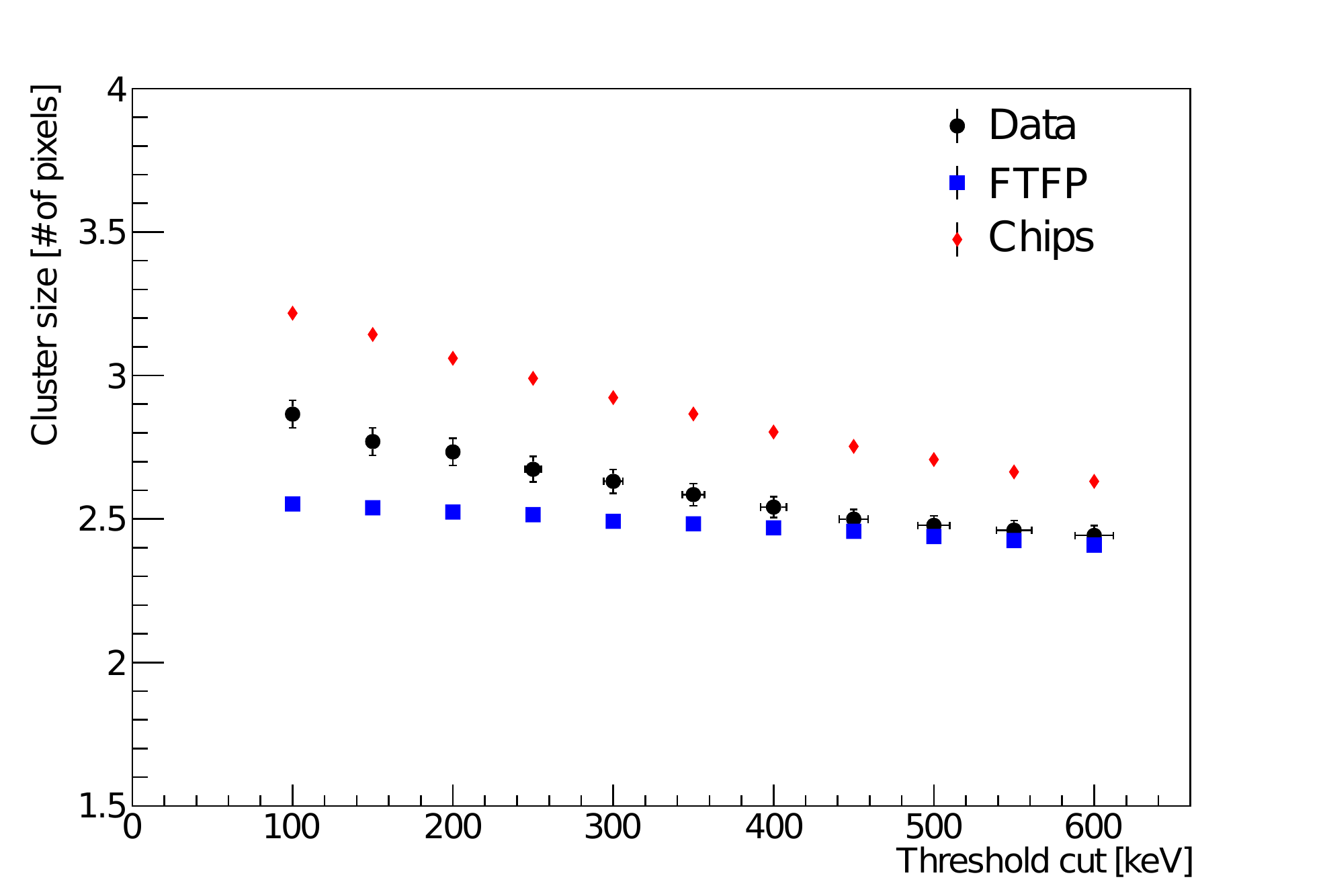}
\caption{Mean cluster size versus the noise cut, excluding the one-pixel clusters.}
\label{fig:cluster_sizegr1}
\end{minipage}%
\hspace*{1cm}
\begin{minipage}{.5\textwidth}
  \centering
  \includegraphics[width=8cm]{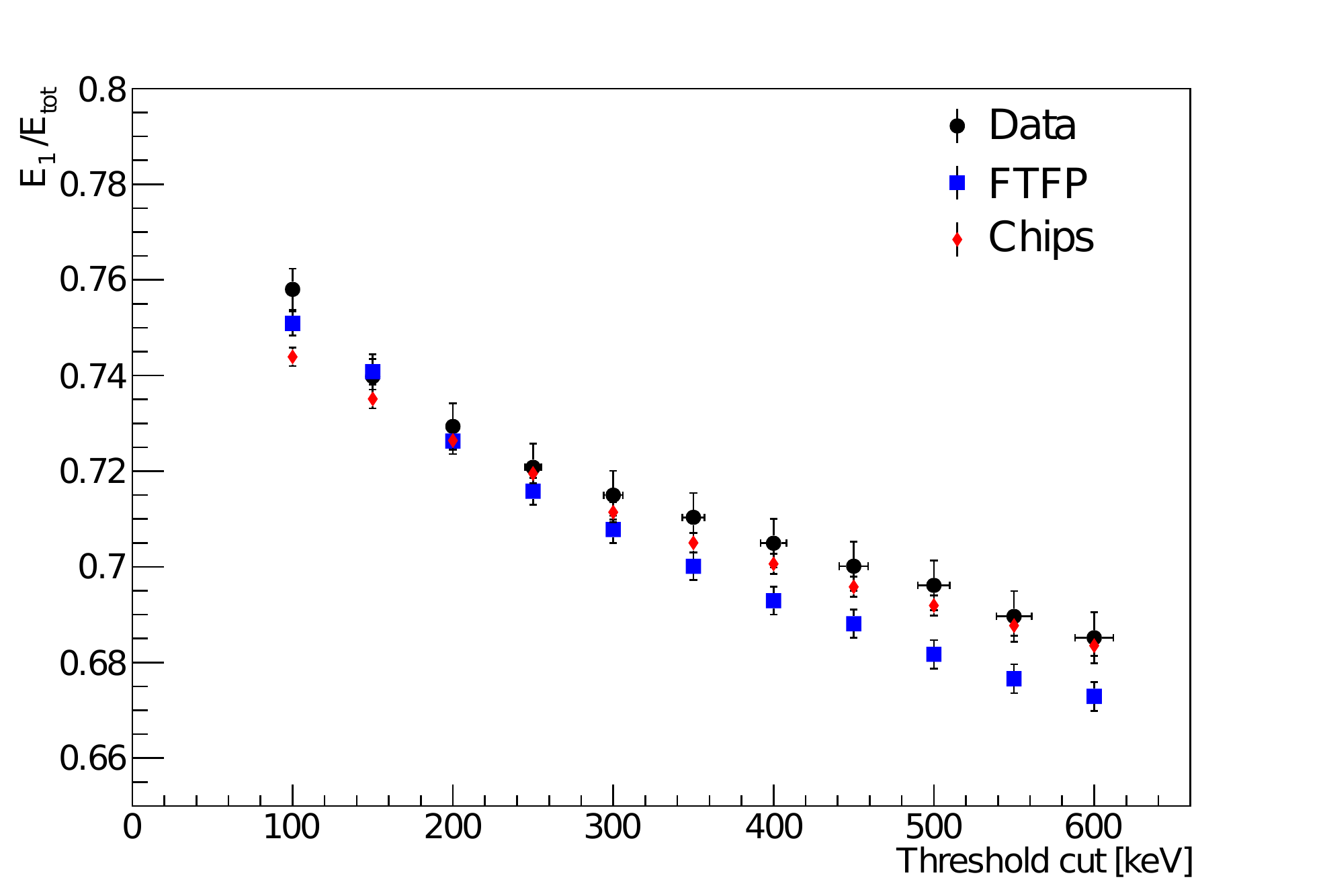}
\caption{$E_1/E_{tot}$-the amount of energy deposited only in the highest energy pixel in the cluster vs the noise cut, excluding the one-pixel clusters.}
\label{fig:E1divEgr1}
\end{minipage}
\end{figure*}

The clustering algorithm developed for the data analysis was also implemented in the simulations. Random gaussian noise was included as well, with the same RMS obtained from the data. As simulations of antiproton annihilations have not yet been validated at low energies, we present a comparison between data and simulations for an energy range of 0-25 MeV of energy released in the detector. Fig. \ref{fig:Etotalgr1} shows  a comparison between data and simulation for the total cluster energy for clusters composed by more than one pixel. Agreement is generally poor with both simulation models up to energies of 5 MeV. At energies above 5 MeV, FTFP shows a better agreement to data. 

Fig. \ref{fig:E1gr1} shows the energy distribution of the highest energy pixel in the clusters for clusters with more than one pixel. When compared with fig. \ref{fig:Etotalgr1}, we see that the total cluster energy distribution is dominated by the contribution of the highest energy pixel. Also in this case the agreement with CHIPS and FTFP in poor $<$ 5 MeV and improve significatively above this energy for the FTFP model. The same validation was made for the quantities $E_2/E_{tot}$ and $E_3/E_{tot}$, showing in this case agreement within statistical errors between data and simulations for both models.

To verify the reliability of the simulations and its dependence on the chosen threshold cut, a scan was performed in the range of 100-600 keV for the same parameters discussed above. The cluster size distribution in fig. \ref{fig:cluster_sizegr1} shows a good description of data points with the FTFP model (with a slight underestimation), while CHIPS systematically overestimates the cluster size to a maximum of $\sim30\%$ at lower cut energies. The relative neutrality of the FTFP is explained with the smaller overall cluster size that the model provides and considering that all the curves tend to the same asymptotic value (2).

The $E_1/E_{tot}$ distribution, with the exclusion of 1 pixel clusters, shows a good agreement between data and FTFP simulations (fig. \ref{fig:E1divEgr1}).
The observed overall negative slope has to be explained with a flattening of the clusters with the increasing cut: clusters having low-energy pixels will be excluded from the statistics. The decreasing $E_1/E$ ratio indicates that the highest pixel energy is not strictly correlated to the total cluster energy.

Good agreement was also found for $E_2/E_{tot}$ and $E_3/E_{tot}$ for both simulation models.
In the case of FTFP, the mean cluster size remains essentially unchanged by the pixel noise cut, while the $E_1/E_{tot}$ ratio shows a strong dependence on the noise cut. 

\section{Summary and Conclusions}

We have successfully measured the first on-sensor annihilations of antiprotons in silicon using a pixelated silicon imaging detector: an important milestone and the first step on the way to designing a novel position sensitive detector for measuring the gravitational effect on antihydrogen. We also performed the first validation of GEANT4 for low energy antiprotons. The main results are as follows:

\begin{itemize}
\item{Study of clusters from antiproton annihilations measuring:}
\begin{itemize}
   \item{Cluster sizes ranging between 1 and 20 pixels, with a mean value of 2.77$\pm0.048$ with the MIMOTERA pixel size ($153\times153\:\mathrm{\mu}$m$^2$, 14 $\mathrm{{\mu}m}$ thickness.)}
  \item{Cluster energies up to 40 MeV.}
  \item{Measurement of prongs up to 2.9 mm.} 
  \item{Discrimination and identification of annihilation products such as protons and heavy ions.}
\end{itemize}

\item{Study of the energy loss of antiprotons in aluminum, validating the simulation with 10$\%$ maximum deviation from experimental data.}

\item{Comparison of two GEANT4 simulation models for low energy antiprotons, CHIPS and FTFP, showing a generally poor agreement for both models at energies $<$5 MeV while FTFP provides a better description of data points for energies $>$5 MeV: while the results are not statistically compatible, the simulation are still providing a reasonable description of the event, especially at higher energies.}
\end{itemize}

These results will allow to identify methods to determine the annihilation position, both by position extrapolation from proton tracks and center of mass methods. It will also be the basis for simulations and design of the first prototype antihydrogen silicon detector for AE$\mathrm{\bar{g}}$IS.

\acknowledgments

We would like to thank the Research Council of Norway and the Bergen Research Foundation for their support. We would like to thank the CERN SSD lab, especially in the persons of Michael Moll and Maurice Glaser, for having provided the instrumentation and infrastructures used for laser calibration measurements. We would also like to thank Alberto Ribon for his help with the fine tuning of GEANT4 with antiprotons as well as with the interpretation of the results.


\clearpage

\begin{thebibliography}{9}
\bibitem{Aegis2007} The AEgIS collaboration - \emph{Proposal for the AEgIS experiment at the CERN antiproton decelerator} 2007
\bibitem{ps200} Los Alamos Report LA-UR 86-260
\bibitem{nieto1991} M. Nieto and T. Goldman \emph{The arguments against antigravity and the gravitational acceleration of antimatter} Physics Reports, vol 205, issue 5, 1991 (221)
\bibitem{Brando1981} T. Brando et al. \emph{Observations of low-energy antineutrons in a time-separated neutral beam} Nuclear Instruments and Methods, vol 180, issues 2-3, 1981 (461)
\bibitem{sn1987a} M. Longo et al. \emph{New Precision Tests of the Einstein Equivalence Principle from SN1987A} - Physical Review Letters, vol 60, no. 3, 1988 (173)
\bibitem{Mills2002} A. P. Mills Jr. and M. Leventhal \emph{Can we measure the gravitational free fall of cold Rydberg state positronium?} Nuclear Instruments and Methods in Phys. Res. B, vol 192, Issues 1-2, 2002 (102)
\bibitem{alphag} The ALPHA Collaboration and A.E. Charman \emph{Description and first application of a new technique to measure the gravitational mass of antihydrogen} - Nature Communications, DOI: 10.1038/ncomms2787
\bibitem{gbar} \emph{The GBAR experiment: gravitational behaviour of antihydrogen at rest} Class. Quantum Grav. 29 184008, 2012
\bibitem{doser2012} M. Doser et al. - \emph{Exploring the WEP with a pulsed cold beam of antihydrogen} Class. Quantum Grav. 29 184009
\bibitem{testera2008} G. Testera et al. \emph{Formation of a cold antihydrogen beam in AEGIS for gravity measurements} AIP Conference Proceedings 1037, 5, 2008 
\bibitem{moire} Markus K. Oberthaler et al. \emph{Inertial sensing with classical atomic beams} Physical Review A, vol 54, 1996 (3165-3176)
\bibitem{Amsler2012} C. Amsler et al. \emph{A new application of emulsions to measure the gravitational force on antihydrogen} Journal of Instrumentation, vol 8, 2013 P02015
\bibitem{Emulsion2013} S. Aghion et al. \emph{Prospects for measuring the gravitational free-fall of antihydrogen with emulsion detectors} Journal of Instrumentation, vol 8, 2013 P08013
\bibitem{McGaughey1986} McGaughey et al. \emph{Low energy antiproton-nucleus annihilation radius selection using an active silicon detector / target} Nuclear Instruments and Methods, vol 249, 1986 (361-365)
\bibitem{Bendiscioli1994}
Bendiscioli G., Kharzeev D. \emph{Antinucleon-Nucleon and Antinucleon-Nucleus Interaction. A Review of Experimental Data} - Rivista del Nuovo Cimento vol. 17, n. 6
\bibitem{StoppingPower} R.Medenwaldt et al. \emph{Measurement of the stopping power of silicon for antiprotons between 0.2 and 3 MeV} Nuclear Instruments and Methods in Phys. Res. B, vol 58, 2002 (1)
\bibitem{Athena2004} The ATHENA collaboration \emph{The ATHENA antihydrogen apparatus} - Nuclear Instruments and Methods in Physics Research A, vol. 518, 2004 (679-711)
\bibitem{Spieler2005} H. Spieler \emph{Semiconductor Detector Systems} - Oxford University Pres, 2005
\bibitem{asterix85} The ASTERIX collaboration \emph{Search for monochromatic pion emission in p$\mathrm{\bar{p}}$ annihilation from atomic p states} Physics Letters B, vol 152, 1985 (135)
\bibitem{SRIM} J.F. Ziegler et al. \emph{SRIM - The Stopping and Range of Ions in Matter} Lulu Press
\bibitem{Hrivnacova} I. Hrivnacova et al. $–$ The Virtual Monte Carlo - Computing in High Energy and Nuclear Physics, La Jolla,
March 24-28, 2003
\bibitem{Degtyarenko} P. V. Degtyarenko, M. V. Kossov, and H.P. Wellisch - \emph{Chiral invariant phase space event generator, I. Nucleon-antinucleon annihilation at rest} - Eur. Phys. J. A 8, 217-222 (2000)
\bibitem{Galoyan} A. Galoyan, V. Uzhinsky \emph{Simulation of Light Antinucleus-Nucleus Interactions} arXiv:1208.3614
\bibitem{Geant4pl} Geant4 Physics Reference Manual
\bibitem{Markiel1988} W. Markiel et al. \emph{Emission of Helium ions after antiproton annihilation in nuclei} - Nuclear Physics A, vol 485, 1988 (445-460)
\bibitem{cnmresistive} D. Bassignana et al. \emph{First investigation of a novel 2D position-sensitive semiconductor detector concept} - Journal of Instrumentation (2012) JINST 7 P02005
\bibitem{Regenfus2003} C. Regenfus \emph{A cryogenic silicon micro-strip and pure-CsI detector for detection of antihydrogen annihilations} - Nuclear Instruments and Methods in Physics Research A 501, 2003, p. 65
\bibitem{Palik85} Edward D. Palik, Handbook of Optical Constants of Solid (1985), Academic Press, NY.
\bibitem{BeamCounter} P. Riedler et al. \emph{Performance of ultra-thin silicon detectors in a 5 MeV antiproton beam} - Nuclear Instruments and Methods in Physics Research A 478, 2002, p.316
\bibitem{Boll2011} R. Boll et al. \emph{Using Monolithic Active Pixel Sensors for fast monitoring of therapeutic hadron beams} - Radiation Measurements vol. 46, Issue 12, 2011 (1971-1973)
\bibitem{Badano2005} L. Badano \emph{D\'{e}veloppement d'un moniteur de faisceau innovant pour la mesure en temps r\'{e}el des faisceau utilis\'{e}s en hadronth\'{e}rapie} - Universit\'{e} Louis Pasteur Strasbourg I (2005)
\bibitem{BollThesis} R. Boll Diploma thesis - University of Heidelberg (2010)








\end{thebibliography}
\end{document}